%% file: asplos24-paper-template.tex
\documentclass[pageno]{jpaper}

\usepackage[normalem]{ulem}
\usepackage{flushend}
\usepackage{tikz}
\usepackage{amsmath}
\usepackage{array}
\newcolumntype{C}[1]{>{\centering\arraybackslash}p{#1}}
\newcolumntype{L}[1]{>{\raggedright\arraybackslash}p{#1}}
\usepackage{caption}
\usepackage{xspace}
\usepackage{graphicx}
\usepackage{booktabs}
\usepackage{pifont}
\newcommand{\cmark}{\ding{51}}%
\newcommand{\xmark}{\ding{55}}%
\usepackage{paralist}
\usepackage{listings}
\usepackage{color}
\definecolor{light-gray}{gray}{0.80}
\definecolor{codegreen}{rgb}{0,0.5,0.0}
\definecolor{codegray}{rgb}{0.5,0.5,0.5}
\definecolor{codepurple}{rgb}{0.58,0,0.82}
\definecolor{backcolour}{rgb}{0.95,0.95,0.92}
\definecolor{codered}{rgb}{1.0,0.01,0.24}
\definecolor{codeblue}{rgb}{0.0, 0.18, 0.65}
\lstdefinestyle{mystyle3}{
    frame=tb,
    commentstyle=\color{codegreen},
    keywordstyle=\color{codered},
    numberstyle=\tiny\color{codegray},
    stringstyle=\color{codepurple},
    basicstyle={\scriptsize\ttfamily},
    breakatwhitespace=false,         
    breaklines=true,                 
    captionpos=b,                    
    keepspaces=true,                 
    numbers=left,                    
    numbersep=5pt,                  
    showspaces=false,                
    showstringspaces=false,
    showtabs=false,                  
    tabsize=2,
    xleftmargin=2ex,
    belowskip=-0.5em,
}

\usepackage{cleveref}
\crefformat{section}{\S#2#1#3} 
\crefformat{subsection}{\S#2#1#3}
\crefformat{subsubsection}{\S#2#1#3}

\usepackage{amsmath}
\usepackage{relsize}

\newcommand{\cc}[1]{\mbox{\smaller[0.5]\texttt{#1}}}

\long\def\ignore#1{}
\newcommand{\mname}{\mbox{RV-CURE}\xspace}
\newcommand{\mech}{\mbox{RV-CURE}\xspace}

\usepackage{url}

\usepackage{breakurl}

\usepackage{makecell}
\usepackage{textcomp}
\usepackage[flushleft]{threeparttable}
\usepackage{enumitem}

\usepackage{cleveref}
\crefformat{section}{\S#2#1#3} 
\crefformat{subsection}{\S#2#1#3}
\crefformat{subsubsection}{\S#2#1#3}
\usepackage{hyperref}

\newcommand*{\fref}[1]{\hyperref[{#1}]{\autoref*{#1}}}
\usepackage{tikz}
\newcommand{\ballwhite}[1]{\tikz[baseline=(myanchor.base)] \node[shape=circle,draw,inner sep=1pt](myanchor) {\bfseries\footnotesize #1};}

\usepackage{algorithm}
\usepackage[noend]{algpseudocode}
\usepackage{multirow}
\usepackage{multicol}
\usepackage{booktabs}
\usepackage{soul}

\begin{document}

\title{RV-CURE: A RISC-V Capability Architecture for Full Memory Safety}

\author{Yonghae Kim\\
\textit{Georgia Institute of Technology}\\
yonghae@gatech.edu
\and
Anurag Kar\\
\textit{Georgia Institute of Technology} \\
anurag.kar@gatech.edu
\and
Jaewon Lee\\
\textit{Georgia Institute of Technology} \\
jaewon.lee@gatech.edu
\and
Jaekyu Lee\\
\textit{Arm Research} \\
jaekyu.lee@arm.com
\and
Hyesoon Kim\\
\textit{Georgia Institute of Technology}\\
hyesoon@cc.gatech.edu
}

\date{}
\maketitle

\input{Tex/A.Abstract}
\input{Tex/B.Introduction}
\input{Tex/C.Background}

\input{Tex/D.ThreatModel}

\input{Tex/E.Overview}
\input{Tex/F.Compiler}

\input{Tex/G.Architecture}
\input{Tex/H.Runtime}
\input{Tex/I.Pruning}
\input{Tex/V.Methodology}
\input{Tex/W.Evaluation}

\input{Tex/X.Security}

\input{Tex/Y.Discussion}
\input{Tex/Y.RelatedWork}
\input{Tex/Z.Conclusion}

\bibliographystyle{plain}
\bibliography{asplos24-paper-template.bib}

\end{document}

%% file: Tex/A.Abstract.tex
\begin{abstract}
Despite decades of efforts to resolve, memory safety violations are still persistent and problematic in modern systems. Various defense mechanisms have been proposed, but their deployment in real systems remains challenging because of performance, security, or compatibility concerns.

In this paper, we propose \mech{}, a RISC-V capability architecture that implements full-system support for full memory safety. For capability enforcement, we first propose a compiler technique, data-pointer tagging (DPT), applicable to protecting all memory types. It inserts a pointer tag in a pointer address and associates that tag with the pointer's capability metadata. DPT enforces a capability check for every memory access by a tagged pointer and thereby prevents illegitimate memory accesses. Furthermore, we investigate and present lightweight hardware extensions for DPT based on the open-source RISC-V BOOM processor. We observe that a capability-execution pipeline can be implemented in parallel with the existing memory-execution pipeline without intrusive modifications. With our seamless hardware integration, we achieve low-cost capability checks transparently performed in hardware. Altogether, we prototype \mech{} as a synthesized RTL processor and conduct full-system evaluations on FPGAs running Linux OS. Our evaluations show that \mech{} achieves strong memory safety at a 10.8\% slowdown across the SPEC 2017 C/C++ workloads.
\end{abstract}

%% file: Tex/B.Introduction.tex
\section{Introduction}

For decades, memory safety issues have exposed substantial security threats to computer systems. Memory safety violations occur when instructions perform illegitimate memory accesses to a program's address space. There are broadly two types of memory safety violations. If a memory access occurs outside of its allowed range, e.g., buffer overflow and out-of-bounds (OOB) access, it violates spatial memory safety. When a memory region is accessed after the region is no longer valid, temporal memory safety is violated, e.g., use-after-free (UAF). Despite long-term efforts to resolve, memory safety issues are still the most prevalent and problematic in the wild. Recent studies~\cite{GoogleReport,BluehatReport} show that $\sim$70\% of security vulnerabilities addressed in industry stemmed from memory violations.

Arguably, the primary hindrance to ensuring memory safety in practice is the performance overhead resulting from defense mechanisms~\cite{SoK_paper}. While various proficient software tools for memory bug detection~\cite{asan,Softbound,BaggyBounds,BaggyBoundsChecking,TaintCheck,EffectiveSan} have been developed, the significant runtime overhead of such tools, e.g., a 73\% slowdown in AddressSanitizer (ASan)~\cite{asan}, has hampered their adoption as a viable runtime solution, limiting their usage to debug and testing.

Hardware-based mechanisms address the performance concern by accelerating security checks with architectural support. Nonetheless, most prior work either requires intrusive hardware changes or lacks security guarantees, impeding their deployment in commodity hardware.

In addition, the history of commercial security solutions reveals that compatibility\footnote{We consider source and binary compatibility (refer to \fref{tab:ptr_comp}).} is one of the major factors hindering their successful deployment~\cite{SoK_paper}. However, maintaining compatibility while ensuring strong memory safety remains challenging, particularly in hardware mechanisms. 

For instance, Intel memory protection extensions (MPX) not only lacked temporal safety but also suffered from compatibility issues as well as a high performance overhead, which caused Intel MPX to be deprecated from the mainstream compiler and OS~\cite{IntelMPX}. In addition, the fat pointer design in the CHERI architectures~\cite{CHERI,Cheri-concentrate,CHERIvoke,Cornucopia} required non-trivial hardware complexity. Despite its strong security guarantees, modifying almost the entire system stacks let the CHERI capability model give up compatibility with legacy code. With such changes, Arm decided to develop the Morello prototype board rather than adding it to their mainstream architectures~\cite{morello}. 

To avoid the non-trivial problems caused by the fat pointer design choice, recent proposals~\cite{AOS,in-fat-pointer,No-FAT,C3} have started investigating a new class of solutions, a pointer-tagging method. This class stores a pointer tag in the unused upper bits of a pointer and uses the tag to look up object metadata or perform an object-based tag match. Such an approach has several advantages. First, it conforms well to conventional programming models since it does not expand the pointer size. Second, metadata propagation overheads can be avoided since pointer tags are propagated along with pointer addresses without explicit register allocations. Third, it enables highly-efficient security checks with the support of a hash table~\cite{AOS}, metadata encoding schemes~\cite{in-fat-pointer}, or pointer/data encryption~\cite{C3}.

Despite such apparent upsides, previous work is shown to struggle in achieving three essential properties together: performance, security, and compatibility. For example, AOS~\cite{AOS} and C\textsuperscript{3}~\cite{C3} only protect heap memory regions, In-Fat Pointer~\cite{in-fat-pointer} does not support temporal safety and No-FAT~\cite{No-FAT} lacks compatibility with legacy code.


To this end, we propose \mech{}, a RISC-V capability architecture that orchestrates compiler, architecture, and system designs for full memory safety. We first propose a pointer-tagging method, \textit{data-pointer tagging} (DPT), applicable to all memory types, including stack, heap, and global memory. Using DPT, \mech{} enforces per-object capability\footnote{E.g., bounds information, privileges, and access permissions.} that defines the legitimacy of memory accesses. Whenever an object is allocated, DPT places a pointer tag in a pointer address and associates that tag with the pointer's capability metadata. To maintain capability metadata, we use a capability metadata table (CMT), which is a software structure stored in memory. During the object's lifetime, DPT validates  every memory access by a tagged pointer via a capability check, thereby preventing any out-of-bounds or non-unprivileged accesses.

DPT also enforces strict temporal safety. Unlike revocation-based approaches requiring non-trivial memory sweep operations~\cite{CHERIvoke,Cornucopia}, DPT lets a tagged pointer (and all its aliases) remain tagged upon object deallocation. DPT instead eliminates its corresponding metadata from the CMT. The use of a dangling (yet tagged) pointer is then destined to trigger a capability check and end up with a failure. 


DPT becomes a generalized approach for all memory types. However, broadening security coverage solely with a compiler technique may result in a non-trivial runtime overhead as shown in many software solutions. To accomplish practical protection, we further investigate efficient, lightweight hardware extensions for DPT. To explore practical design space, we select the RISC-V BOOM core~\cite{BOOM} as our baseline, which is one of the most sophisticated open-source out-of-order processors, and make our hardware changes on top of it.

In our design exploration, we observe that a capability-execution pipeline can be seamlessly integrated with the existing memory-execution pipeline. Given that most security solutions' runtime overhead comes from the memory overhead by security metadata and the execution overhead by security checks, we find that having execution units in the memory pipeline allows us the opportunities of 1) enabling transparent security checks in hardware, 2) utilizing unused cache bandwidth, and 3) enforcing strict hardware-based capabilities.

As we deep-dive into the hardware design, we also identify that hardware resource contentions occurring in the shared data access path become the main performance bottleneck. To mitigate its impact, we further introduce small on-chip buffers, such as a capability cache and store/clear head buffers. We believe such findings in our exploration can commonly apply to modern architectures, e.g., x86, AArch64, and SPARC.

Putting it all together, we prototype \mech{} on top of FireSim~\cite{firesim}, an FPGA-accelerated full-system hardware simulation platform. In our full-system framework, we evaluate the SPEC 2017 C/C++ workloads~\cite{SPEC2017} and show that \mname{} incurs a 10.8\% slowdown on average. To estimate the area and power overheads we make on top of the baseline BOOM core, we synthesize our design using Synopsys Design Compiler using 45nm FreePDK libraries~\cite{FreePDK45} and report 8.6\% area and 11.6\% power overheads. In fact, the BOOM core is (relatively) bare-bones that only implements common architecture components. Hence, we expect a fairly less hardware overhead when assuming the actual deployment in modern processors.
\ignore{
\begin{figure}[t]
\begin{lstlisting}[language=c++, style=mystyle3]
int main(void) {
  char constArr[] = "Hello World";
  char vulArr[10];
  scanf("%s", vulArr);
  printf("constArr: %s\n", constArr);
  printf("vulArr: %s\n", vulArr);
... }
\end{lstlisting}
\caption{A simple buffer overflow example.}
\label{lst:example_code}
\end{figure}
}

In summary, we make the following contributions:

\begin{compactitem}[$\bullet$]
  \item We present \mname{}, a full-system framework that orchestrates compiler, architecture, and system designs.
  \item We propose DPT, a generalized data-pointer tagging method for all memory types protection.
  \item We explore realistic design space on top of the RISC-V BOOM core and realize a synthesizable RTL processor.
  \item \mech{} reports a 10.8\% average slowdown across the SPEC 2017 C/C++ workloads.
\end{compactitem}

%% file: Tex/C.Background.tex
\section{Background}

\subsection{Memory Safety Vulnerability}
Memory safety errors occur when an instruction performs spatially or temporally illegal access to a program's memory address space. Accessing outside its designated memory address range violates spatial memory safety, whereas temporal safety is violated when accessing no-longer-valid memory regions. Despite its notoriety for a long time, recent studies by Microsoft~\cite{BluehatReport} and Google~\cite{GoogleReport} show that memory safety issues are still the most prevalent ($\sim$70\%) and problematic. The resulting side-effects include a system crash, privilege escalation, and information leakage. Given the criticality of such vulnerabilities, ensuring memory safety becomes the task at hand in modern systems.
\ignore{
\fref{lst:example_code} shows a simple yet vulnerable code example where a user provides an input string via \cc{scanf()}. In this example, an attacker can simply inject a long input sequence, whose size exceeds the size of the target array (\cc{vulArr}), to invoke a buffer overflow. Since the unsafe \cc{scanf()} has no knowledge of the array size, it will just store the given input to the target memory address and end up overwriting the adjacent location.
}
\ignore{
\subsection{Challenges in Memory Safety}
\label{ss:challenge}
From previously proposed mechanisms against memory safety vulnerabilities, we identify the following essential but challenging properties to achieve altogether to be an ideal memory protection mechanism. 

\noindent
\textbf{Performance.} Reducing a performance overhead is crucial for a runtime detection solution. For this reason, most software tools tend~\cite{asan,PhASAR,BaggyBounds,Softbound} to be used only for testing and debugging and fail to be deployed at runtime. On the other hand, hardware mechanisms reduce the performance overhead but often compromise security assurance or compatibility.

\noindent
\textbf{Security.} Ensuring strong security guarantees is non-trivial even with a possible trade-off between performance and security. The primary challenge is how to properly maintain security metadata, e.g., memory objects' size and lifetime, and enable spatial and temporal security checks. In particular, many prior studies~\cite{AOS,C3,REST} do not guarantee sub-object safety, i.e., detecting intra-object overflow, while some studies~\cite{Califorms,No-FAT} support that at the cost of painful compatibility loss.

\noindent
\textbf{Compatibility.} Preserving compatibility is important to be deployable in commodity systems. 
Source compatibility is lost if it requires any manual source modification, while binary compatibility is lost if instrumented binaries cannot run with legacy code. The lack of source compatibility may make a solution unscalable for a large software project, and a solution with binary incompatibility cannot run with unmodified user or system libraries.

\noindent
\textbf{Hardware Complexity.} Implementing architectural support for security can bring in a significant performance gain, as evidenced by many prior studies~\cite{Cheri-concentrate,CHERI,CHERIvoke,Watchdog,Hardbound}. However, such hardware schemes often require intrusive hardware changes and discourage CPU vendors from adopting the proposed features. For instance, despite their strong security guarantees, CHERI architectures are known to require non-trivial hardware complexity due to their fat-pointer and co-processor implementations, e.g., 32\% logic overhead~\cite{CHERI}.
}

\subsection{Pointer-Tagging Approaches}
\label{ss:point-tag}

A recent trend shows that pointer-tagging approaches~\cite{No-FAT,C3,AOS,in-fat-pointer,zero21,Califorms} have drawn great attention from research communities. This class places a pointer tag in the upper unused bits of a pointer and uses the tag to look up object metadata in memory or perform an object-based tag match.

As mentioned earlier, such type of mechanisms has shown attractive advantages over previous proposals, e.g., fat-pointer designs~\cite{CHERI,CHERIvoke,CHEx86,Cheri-concentrate,Watchdog,Hardbound}. First, it complies with conventional programming models based on the 64-bit pointer size. Hence, it requires neither physical (or logical) extensions of CPU registers nor additional compiler support for pointer arithmetic, pointer load/store, etc.

Moreover, it enables a zero overhead of metadata propagation since pointer tags are moved along with pointer addresses being stored in upper bits. Note that many previous schemes that maintain per-object metadata have required heavy program instrumentation for explicit metadata propagation for each object, resulting in significant increase in instruction counts and memory overhead. 

With such upsides, prior pointer-tagging schemes have shown promising results. Nevertheless, we still see they have troubles in achieving full memory safety without breaking legacy code compatibility. In the following, we briefly discuss prior work and summarize the comparisons in \fref{tab:ptr_comp}.

\begin{table}[t]
    \centering
    \caption{Comparison between \mech{} and previous schemes.}
    \label{tab:comparison}
    \begin{threeparttable}
    \begin{scriptsize}
    \begin{tabular}{@{}lcccc@{}}
    
    \toprule
    Mechanism & Spatial & Temporal  & Intra-Object & Compatibility Loss~\tnote{1} \\
    \midrule
    AOS~\cite{AOS}                  & \cmark~\tnote{2}     & \cmark~\tnote{2} & \xmark    & $-$                   \\
    C\textsuperscript{3}~\cite{C3}  & \cmark~\tnote{2}     & \cmark~\tnote{2} & \xmark   & $-$                   \\
    In-Fat~\cite{in-fat-pointer}    & \cmark                & \xmark & \cmark                       & $-$     \\
    NoFat~\cite{No-FAT}             & \cmark                & \cmark & \cmark              & Src,Bin    \\
    \midrule
    \mech{}                         & \cmark                & \cmark & \cmark              & $-$           \\
    \bottomrule
    \end{tabular}
    \end{scriptsize}

    \begin{tablenotes}
    \begin{footnotesize}
    \item[1] Source modifications or recompiling the libraries are required, losing source compatibility (Src) or binary compatibility (Bin).
    \item[2] Protection only applies to heap memory regions.

    \end{footnotesize}
    \end{tablenotes}   
\end{threeparttable}
\label{tab:ptr_comp}
\end{table}

AOS~\cite{AOS} proposes a data-pointer-signing scheme that uses the Arm pointer authentication (PA) primitives. Upon \cc{malloc()}, AOS generates a 16-bit pointer authentication code (PAC) and a 2-bit address hashing code (AHC) and associates the PAC with a pointer's bounds metadata. To maintain per-object bounds, AOS uses a bounds-table indexed using PACs with the help of extra hardware extensions. Despite its low-cost operations, its protection is limited to heap memory.

C\textsuperscript{3}~\cite{C3} proposes a stateless mechanism. Based on the power-of-two memory layout, C\textsuperscript{3} assigns each pointer a radix (stored in upper bits) and encrypts the pointer's immutable bit portion. Further, it cryptographically binds pointer encryption to data encryption. Hence, data is stored in memory being encrypted, and only a legitimate pointer can correctly read the data. However, it may allow out-of-bounds accesses within power-of-two slot boundaries. Although a correct read/write is not guaranteed due to encryption schemes, illegal stores could execute to simply corrupt adjacent locations or flip security-critical bits. Also, it provides heap protection only.

To achieve sub-object granular protection, In-Fat Pointer~\cite{in-fat-pointer} presents three metadata encoding schemes, each of which applies to different memory types (i.e., global, stack, heap memory) and a narrowed-bounds retrieval method using an index in a tag. Despite its fine-grained protection, it does not provide temporal safety. Also, its metadata schemes impose several constraints in terms of the number of objects or the maximum size of an object that can be handled.

No-Fat~\cite{No-FAT} observes that the starting address and the size of a pointer can be deduced from a pointer address in the binning allocator. Leveraging the idea, No-Fat extends memory instructions with an extra operand and propagates trusted base addresses along with the corresponding pointers. To verify memory accesses, No-Fat computes the base address of a pointer and compares it to its trusted base. However, to enhance security, it requires instrumenting all stack and global allocations to heap allocation. Furthermore, a source-to-source translation is required for sub-object protection in a way that data layouts of struct types are modified. Also, its mechanism heavily depends on the use of the binning allocator.

\ignore{
\subsection{Prior Work: AOS}\label{ss:aos}
Among the prior work, AOS is prominent for its pointer-signing scheme using the Arm pointer authentication (PA) primitives and its approach to handling hash collisions. As shown in \fref{fig:aos_overview}, AOS generates 16-bit pointer authentication code (PAC) and 2-bit address hashing code (AHC) and then associates the PAC with a pointer's bounds. Since the PAC size is limited, there can exist PAC collisions across different pointers. To deal with the PAC collisions, AOS maintains a multi-way bounds table structure, called a hashed-bounds table (HBT). The HBT is indexed by PACs, and each row can accommodate multiple bounds. To support such a structure, AOS implements an iterative bounds search performed over multiple ways of the HBT.

While AOS presents a promising 8.4\% slowdown, we note that AOS only runs 3-billion instructions using the gem5 simulator~\cite{gem5} for the sake of simulation time. We speculate that running a small portion of programs in AOS might not produce realistic results since memory allocation and access behaviors can vary over the period of program execution. In our evaluations where we entirely execute SPEC CPU 2017 C/C++ workloads, our speculation turns out to be true providing an additional observation.

First, AOS measures the average number of iterative accesses to the HBT to be close to one. However, as shown in Column 3 in \fref{tab:hit_iter}, we observe that the average number of iterations is not trivial even with its proposed bounds way buffer (BWB), 17.03 at a maximum and 3.75 on average. It is obvious that the performance guarantee by AOS would not hold true anymore with such non-trivial iterations. In addition, we observe more frequent bounds-table resizing happen, whereas AOS observes only once for \cc{sphinx3} and twice for \cc{omnetpp}, respectively.

Besides the characteristic of the HBT, AOS only supports heap memory safety. Even though AOS foresees the diminishing number of stack corruption errors, non-trivial error cases on stack memory still exist~\cite{GoogleReport,BluehatReport}. Given that only one memory bug could lead to a chain of other memory violations, complete protection on all memory types is desirable.

\ignore{
While AOS presents promising results, an 8.4\% slowdown, we find two limitations as follows.

[average iteration numbers]
[physicslly-addressed HBT] <<<
However, once we have a virtually-addressed table, new hardware resource contentions.

once we execute entire program, the number of allocation stacks up.

First, AOS only supports heap memory safety. Even though AOS foresees the diminishing number of stack corruption errors, non-trivial error cases on stack memory still exist~\cite{GoogleReport,BluehatReport}. Given that only one memory bug could lead to a chain of other memory violations, complete protection on all memory types is desirable. In addition, AOS does not provide intra-object protection. Second, AOS evaluates its performance using the gem5 simulator~\cite{gem5} by running 3 billion instructions for the sake of simulation time. However, it turns out that running a small portion of a program does not properly capture the overhead of its iterative bounds-search operation. Although AOS reported the average number of iterations close to one, in our evaluation where we run the whole programs, we observe that the average number can significantly increase (up to $\sim$ in SPEC CPU 2017 benchmarks), which would notoriously impact the overall performance.
}

\begin{figure}[t]
\centering
\includegraphics[width=0.8\columnwidth]{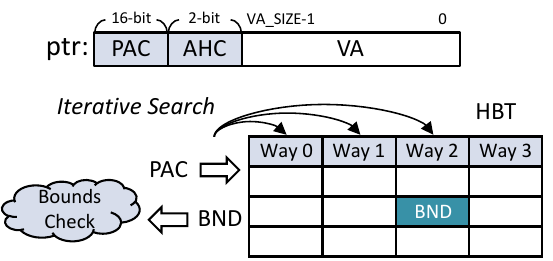}
\caption{Pointer-tagging format and HBT access in AOS.}
\label{fig:aos_overview}
\end{figure}

\begin{figure*}[t]
\centering
\includegraphics[width=2\columnwidth]{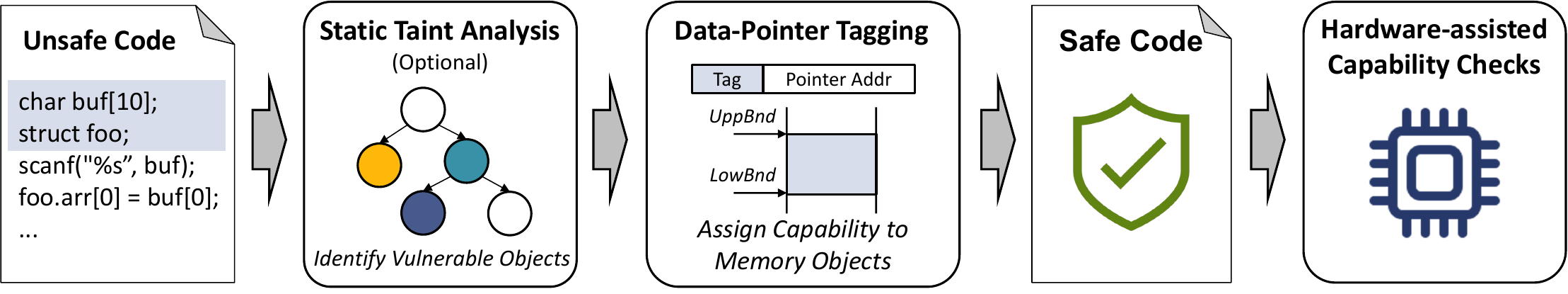}
\caption{Overview of the \mech{} workflow.}
\label{fig:overview}
\end{figure*}

As shown in the above, despite the attractive advantages of pointer-tagging methods, prior work still does not manifest as a complete solution. In addition, a precise performance evaluation using architecture simulators for security solutions requiring metadata management appears to be challenging. From these observations, we bring up two questions; 1) can we generalize a pointer-tagging method to support \textit{full memory safety}? and 2) can we build a \textit{full-stack system} where we can explore practical design space and conduct realistic performance evaluations? In the rest of this paper, we answer these by exploring a full-system framework with a multidisciplinary approach.
}
\ignore{
AOS tags each protected pointer with a 16-bit Pointer Authentication Code (PAC) and uses the pointer tag to index a bounds table in memory to look up the corresponding bounds. To enable bounds checking, AOS implements a memory check unit (MCU), as shown in \fref{fig:hardware}, that takes memory access instructions, i.e., load and store, as well as bounds instructions, i.e., \cc{bndstr} and \cc{bndclr}. Depending on the bounds-instruction type, the MCU generates memory requests to store or clear bounds in a hashed bounds table (HBT). Since the HBT is indexed by PACs (embedded in memory addresses), the locations of bounds are calculated using the base address of the HBT and PACs, i.e., \cc{HBT[PAC]}. The MCU examines addresses of memory instructions to determine the necessity of bounds checking. If an address embedding a nonzero PAC is detected, the MCU loads its corresponding bounds and validates the access via hardware-based bounds checking. AOS extends the instruction set architecture (ISA) for the required operations.
}

\ignore{
\begin{compactitem}[$\bullet$]
    \item \textbf{\cc{pacma ptr, mod}}. The \cc{pacma} is used to \textit{sign} a pointer. Using the QARMA block cipher~\cite{QARMA}, it calculates a PAC using a pointer address (\cc{ptr}), a modifier~\footnote{A modifier is a 64-bit tweak to differentiate contexts of a program.} (\cc{mod}), and a PA key.\footnote{New PA keys invisible to user are assigned to each new process.} The PAC is stored in the upper bits of the \cc{ptr}.
    
    \item \textbf{\cc{xpacm ptr}}. The \cc{xpacm} \textit{strips} a pointer to remove the PAC located in the upper bits of the pointer (\cc{ptr}).
    
    \item \textbf{\cc{bndstr ptr, size}}. The \cc{bndstr} calculates bounds using a pointer (\cc{ptr}) and a size and stores the bounds in the location indexed by the PAC of the \cc{ptr} in an HBT.
    
    \item \textbf{\cc{bndclr ptr}}. The \cc{bndclr} clears bounds from the location indexed by the PAC of a pointer (\cc{ptr}) in an HBT.
\end{compactitem}

Since the PAC has a limited size, some pointers can share the same PAC, resulting in PAC collisions. Consequently, the HBT is implemented as a multi-way structure to handle the collisions by accommodating multiple bounds for the same PAC. While such a design makes the HBT scalable, bounds instructions will perform an iterative bounds search over multiple ways. To lower the overhead, a bounds way buffer (BWB) keeps track of recent bounds-table accesses and tries to give correct locations for subsequent accesses upon their first attempts.
}

\ignore{
\subsection{Challenges in Memory Safety}
\label{ss:challenge}

Numerous prior proposals have been presented to prevent memory safety vulnerabilities. While we defer the discussion on related work to \fref{s:related_work}, we identify three major challenges of hardware-based mechanisms in achieving robust (security coverage), efficient (performance), and deployable (compatibility) memory safety.

\noindent
\textbf{How to support full security coverage.}
The primary challenges in building a hardware-based framework are: 1) how to properly expose software semantics to hardware and 2) how to enable efficient hardware operations using such information. Specifically, we should expose each memory object's size and lifetime to hardware and implement an efficient detection mechanism for illegal accesses. To protect the entire memory space, the distinct allocation method and lifetime of each memory type should be carefully addressed. We also need to devise how to maintain metadata and enforce fine-grained safety at the sub-object granularity. 

\noindent
\textbf{How to minimize slowdown by security checks.}
To enhance the security level of systems, security-check statements are typically defined and inserted in programs based on the targeted threat model and security coverage. However, given that security checks are always needless, in terms of the correctness of program execution, excessive check operations can cause a significant slowdown, which is unacceptable in most cases. To minimize the slowdown, it is important to define a threat model built upon sophisticated attack models and apply a minimally required set of security checks to sustain performance but not hamper security.

\noindent
\textbf{How to retain compatibility with legacy code.}
Preserving compatibility is important to be practical and deployable in commodity systems. For example, a solution loses source compatibility if it requires any manual source modification. Even minimal human intervention, such as code annotation and source-to-source translation,  may make a solution unscalable and impractical for a large software project. Binary compatibility is lost if instrumented binaries cannot run with legacy code, e.g., unmodified libraries. Given that system libraries are typically protected and unchangeable without high privileges, the binary incompatibility not only prevents incremental development but also limits the operating platforms.
}

%% file: Tex/D.ThreatModel.tex
\section{Threat Model}
\label{s:threat}
In our threat model, attackers can exploit one or more memory bugs that may exist in the address space of a user process and thereby obtain arbitrary read and write capabilities to leak sensitive information or gain root permissions via privilege escalation. We do not limit the scope of memory regions. Therefore, memory bugs can exist anywhere in a program, including stack, heap, and global memory regions. 

In past few years, speculative execution attacks~\cite{spectre,meltdown,foreshadow,zombieload} have revealed significant threats.
Such attacks may be considered of a different type from memory safety, since those are hardware-oriented (i.e., derived from out-of-order execution in modern CPUs) whereas memory safety issues are software bugs. Also, memory safety bugs are the result of correct execution unlike that of speculative attacks. Hence, typical mitigations for the two tend to be orthogonal, although No-Fat~\cite{No-FAT} presented using memory safety defense for Spectre-V1 resiliency. As such, we leave out of scope such hardware-oriented attacks, including side-channel~\cite{primeprobe,flushreload,eddie,power_channel} and row-hammer attacks~\cite{row_hammer,RAMBleed,ECCploit,PThammer}.

Meanwhile, the recent disclosure of PACMAN~\cite{pacman} showed that memory protection schemes can be defeated with a synergy of memory corruption vulnerability and speculative execution attacks. Given that only a correct PAC guess can output a valid pointer, PACMAN observed micro-architectural side effects of PA and was able to leak PACs  without causing crashes. PACMAN could break a common assumption that randomized tags are effective in protecting pointers, believed in most schemes utilizing the high-order bits of a pointer.

However, we believe that such findings retroactively emphasize the need for stronger memory safety solutions. In fact, the PACMAN attack needed to trigger a true buffer overflow to proceed with a control-flow attack, and Arm PA alone could not prevent it. If a stronger mechanism was in effect, the attack could not have proceeded. In this sense, \mech{} does not prevent the PACMAN attack, but can operate as a defense that ceases attack transitions from memory safety bugs.


%% file: Tex/E.Overview.tex
\section{Overview}
\label{s:overview}

In this section, we overview \mech{}, a full-system architecture for memory safety. In achieving memory safety, we identify four properties, essential but challenging to achieve altogether. We introduce our approach to each as follows.

\noindent
\textbf{Security.} Prior work tends to specialize in specific memory type protection, e.g., heap protection~\cite{AOS,C3,CHEx86}. To overcome this limitation, we propose a generalized method for all memory types, called \textit{data-pointer tagging} (DPT) (\cref{ss:ensuring}).



\noindent
\textbf{Compatibility.} Preserving compatibility is crucial to be deployable in commodity systems. By taking a pointer-tagging approach, we conform to conventional programming models. Hence, instrumented programs can run with unmodified legacy code, i.e., binary compatibility is retained. Moreover, our compiler support requires neither source code modifications nor data structure changes. Thus, source compatibility is assured.


\noindent
\textbf{Hardware \& Performance.} Despite a number of novel architecture proposals, intrusive hardware changes have often discouraged CPU vendors from adopting new designs. For this reason, based on the BOOM core~\cite{BOOM}, we explore realistic design space and propose practical hardware extensions (\cref{s:hardware}). 

\ignore{
\noindent
\textbf{Achieving full memory safety.} We propose \textit{data-pointer tagging (DPT)} that generates a 16-bit cyclic redundancy check hash (CRC-16), used as a pointer tag, and stores the hash in the high-order bits of a pointer. As mentioned in \fref{ss:point-tag}, the pointer-tagging idea itself is not new. However, prior work tends to specialize in specific memory type protection, e.g., heap protection~\cite{AOS,C3}. In this work, we investigate a generalized approach that is applicable to protecting all memory types and ensuring fine-grained sub-object safety in synergy with our custom hardware support.


\noindent
\textbf{Maintaining compatibility with legacy code.} Since we utilize the unused bits of a pointer, unlike fat pointer designs with a modified pointer size~\cite{CHERI,CHERIvoke,Cheri-concentrate,Cornucopia}, we conform to conventional language models. Hence, the instrumented programs can run with unmodified legacy code. Also, our compiler passes are designed at the LLVM intermediate representation (IR) level without source modifications. Therefore, we retain both source and binary compatibility.

\noindent
\textbf{Pruning unnecessary security checks.} Extending security coverage results in performance loss due to the increased security metadata and checks. To optimize performance, we rely on the known security concept that memory safety vulnerabilities are typically exploited through untrusted external inputs~\cite{TaintCheck,scrash,perl}. To leverage this concept, we design static taint analysis, which helps identify vulnerable objects in a program and avoid unnecessary operations for untainted (safe) objects. We present more details in \fref{ss:taint}.
}



\ignore{
\subsection{Design Changes in \mech{}}
\label{ss:hardware_extension}

Compared to AOS, we make the following design changes in \mech{}. First, we create an 18-bit PAC as a pointer tag, whereas AOS generates 16-bit PAC and 2-bit AHC. In AOS, AHC (Address Hashing Code) is used to calculate an address hash used to look-up the BWB and to determine if a pointer has been signed. Note that while AOS's protection is limited to heap regions, \mech{} protects all memory types. Hence, \mech{} needs to sign and manage more data pointers, leading to more PAC collisions. Our empirical studies find that the benefit of using longer PACs is greater than that of using AHCs when non-trivial PAC collisions occur. Without using AHCs, \mname{} ensures that the \cc{pacma} always generates a nonzero PAC and determines a pointer's \textit{signess} by checking if its PAC is zero or not. Second, we introduce a new instruction, \cc{bndsrch}, to enable practical sub-object safety. In \fref{sss:protection}, we present the details on when and how the \cc{bndsrch} is used.

\vspace{1pt}
\begin{compactitem}[$\bullet$]
    \item \textbf{\cc{bndsrch ptr, size}}. The \cc{bndsrch} calculates bounds using a pointer address (\cc{ptr}) and a size and then finds the same bounds in the location indexed by the PAC of the \cc{ptr} in an HBT. If found, it returns 1 (\cc{true}), and otherwise 0 (\cc{false}).
\end{compactitem}
\vspace{1pt}
}

%% file: Tex/F.Compiler.tex
\section{Compiler Support}
\label{ss:ensuring}

In this section, we introduce new instruction-set architecture (ISA) extensions and present the proposed pointer-tagging method enabled with new instruction types. To implement DPT, we design custom passes in the LLVM framework~\cite{LLVM}.

\ignore{
\subsection{Object Allocation Methods and Lifetime}
\label{sss:allocation}

\noindent
\textbf{Stack object.} With type information, a stack object is declared using an \cc{alloca} instruction. Its memory region is allocated on the stack frame upon the function call and is released when the function returns to its caller.

\noindent
\textbf{Heap object.} With a size argument, heap objects are declared using dynamic allocation functions, such as \cc{malloc()}. In LLVM IR, the integer pointer type (\cc{i8*}) is always returned from an allocation function and is cast to the object type to use. The allocated objects are valid until being freed using library functions, such as \cc{free()}.

\noindent
\textbf{Global object.} With a \cc{global} or \cc{constant} qualifier, global objects are declared with an explicit type in the global scope. Once a program starts, those are initialized and alive until the program ends.

Despite various allocation methods, we note that the size of each object is explicitly provided; for stack and global objects, the size can be calculated using the type information based on a given data layout. This indicates that LLVM IR preserves enough information to maintain the exact bounds of memory objects of any type.
}

\begin{figure}[t]
\centering
\includegraphics[width=0.9\columnwidth]{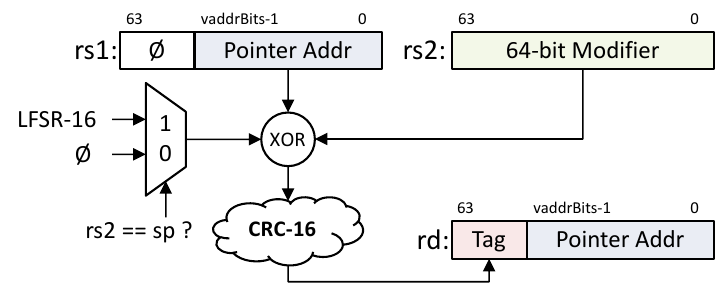}
\caption{Data-pointer tagging using a \cc{tagd} instruction.}
\label{fig:tagd}
\end{figure}

\subsection{ISA Extensions} \label{ss:isa_extensions}
To allow for DPT, we introduce new instructions in the RV64I base integer instruction set~\cite{RISC-V-ISA}.


\noindent
\textbf{\cc{tagd rd, rs1, rs2}}: takes as input a pointer address (\cc{rs1}) and a modifier (\cc{rs2}). As shown in \fref{fig:tagd}, it may use an additional value (\cc{LFSR-16}) that is XOR-ed with \cc{rs1} and \cc{rs2} if \cc{rs2} is the stack pointer (\cc{sp}). The XOR-ed value is then used to calcuate a CRC-16 hash.\footnote{We use CRC-16 for tag generation since we hope our design to be freely used in any commercial or open-source projtects. Other cryptographic algorithms~\cite{QARMA,k-cipher} can be used instead.} It returns a tagged pointer address (\cc{rd}) in which the generated hash is stored in the high-order bits. \cc{LFSR-16} refers to 16 bits returned from a linear-feedback shift register (LFSR) that generates pseudo-random numbers. We find using it is effective in making tags unpredictable and also mitigating use-after-reallocation attacks (see \fref{sec:security}).


\noindent
\textbf{\cc{xtag rd, rs1}}: takes as input a tagged pointer address (\cc{rs1}) and zeros its tag. It returns a stripped pointer address (\cc{rd}).
    
\noindent
\textbf{\cc{cstr rs1, rs2}}: takes a tagged pointer address (\cc{rs1}) and an object size for bounds calculation (\cc{rs2}). It calculates capability metadata and stores in a capability metadata table (CMT).

\noindent
\textbf{\cc{cclr rs1}}: takes a tagged pointer address (\cc{rs1}) and clears its corresponding capability metadata from a CMT.

\ignore{
\begin{compactitem}[$\bullet$]
    \item \textbf{\cc{tagd rd, rs1}}: takes as input a pointer address (\cc{rs1}) and and calculates a CRC-16 hash using the given input. It places the generated hash, used as a pointer tag, in the upper bits of the pointer and returns a tagged pointer address (\cc{rd}).
    
    
    \item \textbf{\cc{xtag rd, rs1}}: takes as input a tagged pointer address (\cc{rs1}) and zeros its tag. It returns a stripped pointer address (\cc{rd}).
    
    
    \item \textbf{\cc{cstr rs1, rs2}}: takes as input a tagged pointer address (\cc{rs1}) and a size (\cc{rs}). It calculates capability metadata for the pointer and stores it in a capability metadata table (CMT).
    
    
    \item \textbf{\cc{cclr rs1}}: takes a tagged pointer address (\cc{rs1}) and clears its corresponding capability metadata stored in a CMT.
    \ignore{
    \item \textbf{\cc{csrch rs1, rs2}}: takes a tagged pointer address (\cc{rs1}) and a size (\cc{rs2}). It calculates capability metadata using the given inputs and searches for the metadata in a CMT. If the same metadata are found, it returns one, otherwise zero.
    }
\end{compactitem}
}

In \mech{}, capability metadata may hold memory capabilities, such as bounds information and read/write/execute permissions. However, we note that the baseline BOOM core has 8B cache-access bandwidth, which limits the metadata size to 8 bytes that can be loaded by a load. To use a single memory request for loading metadata, we use an 8B bounds encryption scheme proposed by Kim et al.~\cite{AOS}. Note that modern processors support up to 64B cache access bandwidth. Hence, once wider bandwidth has been supported, more various memory capabilities can be enforced using the same mechanism.



\begin{figure}[t]
\centering
\includegraphics[width=0.8\columnwidth]{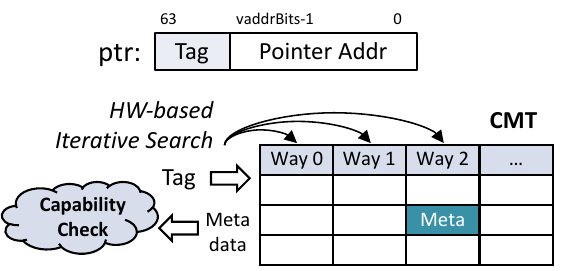}
\caption{Capability metadata access to the CMT.}
\label{fig:CMT}
\end{figure}

\begin{figure}[t]
\begin{lstlisting}[language=c++, style=mystyle3, escapechar=!]
  char stackObj[SIZE];        // Alloc stack obj
!\colorbox{gray!20}{\ \ \color{codered}{char} \color{black}{*shadowPtr = stackObj;}\ \ \ \ \ \ \ \ \ \ \ \ \ \ \ \ \ \ \ \ \ \ \ \ \ \ \ \ \ \ \ \ \ \ }!
!\colorbox{gray!20}{\ \ \color{codeblue}{tagd} \color{black}{shadowPtr, sp;}\ \ \ \ \ \ \ \ \ \ \ \ \ \ \ \ \ \ \ \ \ \ \ \ \ \ \ \ \ \ \ \ \ \ \ \ \ \ \ \ \ \ }!
!\colorbox{gray!20}{\ \ \color{codeblue}{cstr} \color{black}{shadowPtr, SIZE;}\ \ \ \ \ \ \ \ \ \ \ \color{codegreen}{// Store capability}\ \ \ \ \ \ \ \ \ \ }!
!\colorbox{gray!20}{\ \  shadowPtr[0] = 'a';\ \ \ \ \ \ \ \ \ \ \ \ \ \color{codegreen}{// stackObj[0] = 'a';}\ \ \ \ \ \ \ \ }!
!\colorbox{gray!20}{\ \ \color{codeblue}{cclr} \color{black}{shadowPtr;} \ \ \ \ \ \ \ \ \ \ \ \ \ \ \ \ \color{codegreen}{// Release capability} \ \ \ \ \ \ \ }!
\end{lstlisting}

\vspace{0.2cm}
\centering \small (a) Stack object protection.

\begin{lstlisting}[language=c++, style=mystyle3, escapechar=!]
  char *heapPtr=(char*)malloc(SIZE); // Alloc heap obj
!\colorbox{gray!20}{\ \ \color{codeblue}{tagd} \color{black}{heapPtr, sp;}\ \ \ \ \ \ \ \ \ \ \ \ \ \ \ \ \ \ \ \ \ \ \ \ \ \ \ \ \ \ \ \ \ \ \ \ \ \ \ \ \ \ \ \ }!
!\colorbox{gray!20}{\ \ \color{codeblue}{cstr} \color{black}{heapPtr, SIZE;}\ \ \ \ \ \ \ \ \ \ \ \ \ \color{codegreen}{// Store capability}\ \ \ \ \ \ \ \ \ \ }!
  heapPtr[0] = !\color{black}{'a'}!;
!\colorbox{gray!20}{\ \ \color{codeblue}{cclr} \color{black}{heapPtr;}\ \  \ \ \ \ \ \ \ \ \ \ \ \ \ \ \ \ \ \color{codegreen}{// Release capability} \ \ \ \ \ \ \ }!
!\colorbox{gray!20}{\ \ \color{codeblue}{xtag} \color{black}{heapPtr;}\ \ \ \ \ \ \ \ \ \ \ \ \ \ \ \ \ \ \ \ \ \ \ \ \ \ \ \ \ \ \ \ \ \ \ \ \ \ \ \ \ \ \ \ \ \ \ \ }!
  free(heapPtr);
\end{lstlisting}

\vspace{0.2cm}
\centering \small (b) Heap object protection.
\label{lst:heap}

\begin{lstlisting}[language=c++, style=mystyle3, escapechar=!]
char glbObj[SIZE];             // Alloc global obj
void foo() {
!\colorbox{gray!20}{\ \ \color{codered}{char} \color{black}{*shadowPtr = \&glbObj;}\ \ \ \ \ \ \ \ \ \ \ \ \ \ \ \ \ \ \ \ \ \ \ \ \ \ \ \ \ \ \ \ \ \ \ }!
!\colorbox{gray!20}{\ \ \color{codeblue}{tagd} \color{black}{shadowPtr, TYPE;}\ \ \ \ \ \ \ \ \ \ \ \ \ \ \ \ \ \ \ \ \ \ \ \ \ \ \ \ \ \ \ \ \ \ \ \ \ \ \ \ }!
!\colorbox{gray!20}{\ \ shadowPtr[0] = 'a';\ \ \ \ \ \ \ \ \ \ \ \ \ \ \color{codegreen}{// glbObj[0] = 'a';}\ \ \ \ \ \ \ \ \ }!
}
int main(void) {
!\colorbox{gray!20}{\ \ \color{codered}{char} \color{black}{*shadowPtr = \&glbObj;}\ \ \ \ \ \ \ \ \ \ \ \ \ \ \ \ \ \ \ \ \ \ \ \ \ \ \ \ \ \ \ \ \ \ \ }!
!\colorbox{gray!20}{\ \ \color{codeblue}{tagd} \color{black}{shadowPtr, TYPE;}\ \ \ \ \ \ \ \ \ \ \ \ \ \ \ \ \ \ \ \ \ \ \ \ \ \ \ \ \ \ \ \ \ \ \ \ \ \ \ \ }!
!\colorbox{gray!20}{\ \ \color{codeblue}{cstr} \color{black}{shadowPtr, SIZE;} ...\ \ \ \ \ \ \ \ \color{codegreen}{// Store capability}\ \ \ \ \ \ \ \ \ }!
\end{lstlisting}

\vspace{0.2cm}
\centering \small (c) Global object protection.
\label{lst:global}

\begin{lstlisting}[language=c++, style=mystyle3, escapechar=!]
struct S {
  int num;
  char subObj[10];
};
int foo(void) {
  char *objPtr = (struct S*) malloc(SIZE); // Alloc str
  // Sub-object indexing: GEP(objPtr, 0, 1)
!\colorbox{gray!20}{\ \ \color{codered}{char} \color{black}{*subObjPtr = }\color{codeblue}{mutate\_ptr}\color{black}{(objPtr->subObj, SIZE);}\ \ \ \ \ \ \ \ \ \ }!
!\colorbox{gray!20}{\ \  subObjPtr[0] = 'a'; ...\ \ \ \ \ \ \ \ \ \ \color{codegreen}{// objPtr->subObj[0] = 'a';}\ }!
\end{lstlisting}

\vspace{0.2cm}
\centering \small (d) Intra-object protection.

\caption{Memory object protection using DPT.}
\label{fig:dps}
\end{figure}

\begin{figure}[t]
\begin{lstlisting}[language=c++, style=mystyle3, escapechar=!]
// 1st and 2nd-level card tables, respectively
char *cardTable, *cardTable2nd; 
void *mutate_ptr(void *ptr, size_t size) {
  !\color{codeblue}{tagd}! ptr, TYPE;
  size_t idx = (ptr >> 3);
  char bitmask = (0x1 << (ptr & 0x7));
  char mark = (cardTable[idx] & bitmask);
  if (mark == 0) {
    !\color{codeblue}{cstr}! ptr, size;
    cardTable[idx] |= bitmask;
    size_t idx2 = (ptr >> 9);
    cardTable2nd[idx2] = 1;
  }
  return ptr; }
\end{lstlisting}
\caption{Dynamic pointer mutation with card marking.}
\label{fig:mutate}
\end{figure}

\begin{figure*}[t]
\centering
\includegraphics[width=2.1\columnwidth]{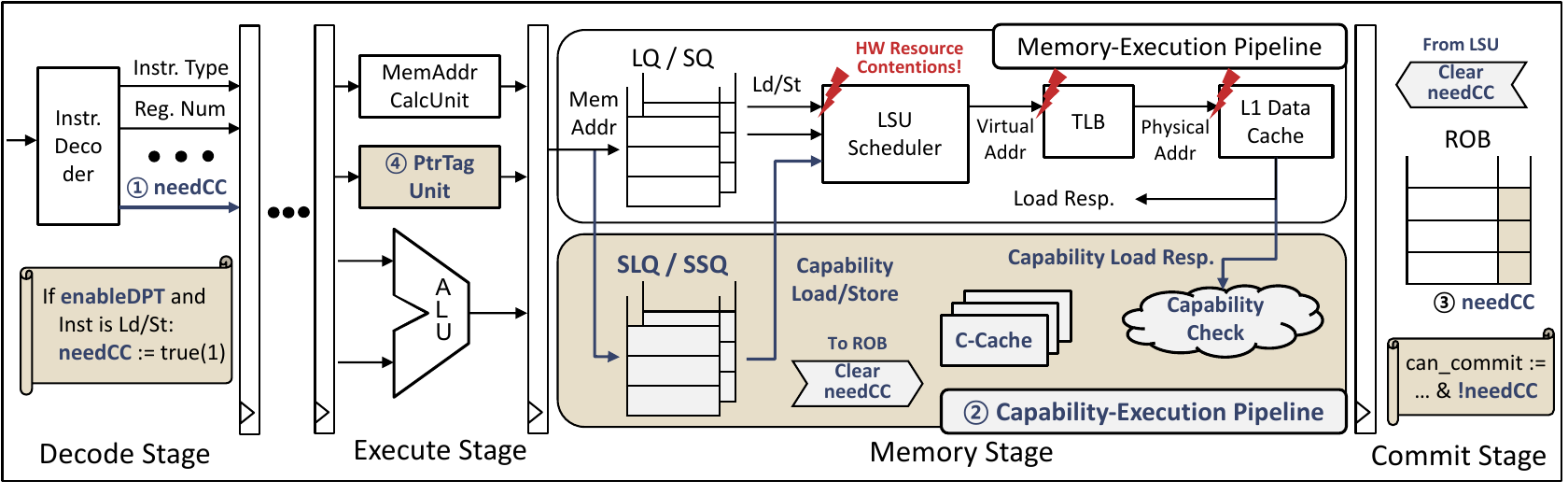}
\caption{Overview of the hardware modifications in \mech{}.}
\label{fig:rv_cure}
\end{figure*}

\subsection{Data-Pointer Tagging (DPT)}
\label{sss:protection}
With the extended ISA, we develop DPT that assigns capability metadata upon object allocation and releases the assigned metadata upon deallocation. Capability metadata are maintained in the CMT, allocated in memory at program entry. The basic mechanism of DPT is to enforce capability checks during an object's lifetime. Whenever a memory access by a tagged pointer is detected, the access is validated with its corresponding metadata. DPT also ensures strict temporal safety. When an object is deallocated, DPT eliminates its metadata from the CMT and lets its pointer (including its aliases) remain tagged. All subsequent accesses using dangling (yet tagged) pointers are destined to trigger and fail in capability checks.

To associate capability metadata, DPT tags each pointer using a \cc{tagd} instruction that embeds its pointer tag in the pointer address. DPT then uses a \cc{cstr} instruction to calculate and store its capability metadata in the CMT. The location of capability metadata is computed using the base address of the CMT and a pointer tag, i.e., \cc{CMT[tag]}. Upon object deallocation, DPT clears its corresponding metadata from the CMT using a \cc{cclr} instruction. A \cc{xtag} instruction can be used when a pointer tag needs to be zero-ed out.

For efficient metadata management, we adopt a set associative metadata table from prior work~\cite{AOS}, which we call the CMT in our work. Given the limited tag size (we use 16 bits), tag collisions can happen across pointers sharing the same tag. The multi-way structure of the CMT deals with the issue by accommodating multiple sets of metadata for each tag. With such a structure, as shown in \fref{fig:CMT}, we implement an iterative search mechanism that allows efficient metadata accesses over multiple ways. In our evaluations, we start with the CMT with four ways, of which size is 2MB (=8B*4*2\textsuperscript{16}), and we observe the initial size is large enough for most applications.


Next, we detail our approach for DPT as follows. 

\noindent
\textbf{Stack object protection.}
Stack objects may not have a pointer pointing to themselves whereas heap objects always have explicit pointers. To protect a stack object, we create a shadow pointer and associate it with the object's capability metadata (see lines 3-4 in \fref{fig:dps}a). Then, we have any memory access to the object done via the tagged pointer's dereferencing (see line 5 in \fref{fig:dps}a). The capability metadata are released right before the current function returns (see line 6 in \fref{fig:dps}a). 


\noindent
\textbf{Heap object protection.}
Dynamic memory allocation methods always return an integer pointer type (\cc{i8*}). Hence, we tag a returned pointer and store its capability metadata right after the allocation (see lines 2-3 in \fref{fig:dps}b). Its metadata are released right before the heap object is deallocated (see lines 5-6 in \fref{fig:dps}b). The \cc{xtag} is inserted to remove the pointer's tag to avoid capability checks during \cc{free()}.


\noindent
\textbf{Global object protection.}
Like stack objects, global objects may not have an explicit pointer. Hence, in a program entry, e.g., \cc{main()}, we first create a (local) shadow pointer and associate each global object's capability metadata (see lines 8-10 in \fref{fig:dps}c). Since global objects are accessible anywhere, we also create shadow pointers in each function where global objects are used (see lines 3-4 in \fref{fig:dps}c). Note that we store the metadata only once in the program entry to avoid making metadata duplicates. We do not release metadata because global objects are alive until a program finishes.

When tagging stack and heap object pointers, we use a stack pointer (\cc{sp}) as a modifier, which is XOR-ed with \cc{LFSR-16} in hardware. For global object pointers, we use an encoded 64-bit type identifier used in prior work~\cite{pacitup}, i.e., \cc{tagd ptr, TYPE}, to obtain the same tag for the same pointer anywhere.

\subsection{Fine-grained Protection using DPT}
As explained, DPT manifests as a common approach to all memory types. In this section, we show that DPT can be further generalized for sub-object safety, i.e., protection against intra-object overflows (e.g., overflows between struct fields). 

Our primary intuition is that if we create a new pair of a pointer tag and narrowed bounds for a sub-object and associate the pair with the sub-object, we could enforce fine-grained checks.
We note that a \cc{GEP} instruction is used to index a sub-object from an object in LLVM IR. This allows us to detect each sub-object indexing. However, the \cc{GEP} calculates the sub-object address simply by adding an offset to its base address, and hence the sub-object inherits the same pointer tag from the parent object. To differentiate sub-objects, we choose to re-tag (mutate) pointers returned by \cc{GEP}s to assign different tags, represented as  \cc{mutate\_ptr()} in line 8 in \fref{fig:dps}d. We call this method \textit{dynamic pointer mutation}.


However, we find one challenge regarding when to store sub-object's narrowed metadata. One naive approach would be storing those after each GEP. However, this may create numerous duplicates of the same metadata. This implies the need for keeping track of past sub-object metadata allocations.

To this end, motivated by a card marking mechanism~\cite{writebarrier} used in tracing garbage collectors~\cite{g1,shenandoahgc,zgc}, we adopt a two-level card table design. Similar to shadow memory used in many software solutions~\cite{asan,valgrind}, the first-level table reserves 1/8 of the virtual address space and maps each byte to one bit in memory space, and the second-level one reserves 1/64 of the virtual space and maps 512 bytes to one byte. The complete operations of \cc{mutate\_ptr()} are presented in \fref{fig:mutate}.

When sub-object indexing occurs, the sub-object pointer is re-tagged using a \cc{tagd}. We use a 64-bit type identifier of \cc{i8*} as a modifier here. Then, the first-level card table is looked up using the sub-object base address. If the corresponding mark bit is not dirty, it means its metadata have not been stored. Hence, the metadata are newly stored using a \cc{cstr}, and the corresponding bits of the two card tables are marked dirty. Otherwise, \cc{cstr} and card marking operations are skipped.

Additionally, upon deallocation, we require the task of scanning card table entries corresponding to an object's memory region in order to eliminate its all sub-object metadata. This time, the second-level card table entry is first looked up for marking information at the 512B granularity. If its mark bit is dirty, then the corresponding entries of the first-level table are checked. The scan operation is only required for struct-type stack objects and all heap objects, since the type of heap objects is not known at \cc{free()}.



\ignore{
\noindent
\textbf{Intra-object protection.} Typical GCs implement a linked list to maintain memory reference information. However, in our evaluation, such an approach is shown to impose an intolerable slowdown (up to $\sim$3x) as typical GCs do. Not to be overwhelmed by the overhead, we propose the use of a new \cc{bndsrch} instruction (see \fref{ss:hardware_extension}). Having a similar mechanism as the \cc{bndstr} and \cc{bndclr}, the \cc{bndsrch} takes the address and size of an indexed sub-object, accesses the location indexed by the mutated pointer's PAC in the HBT, and efficiently checks the existence of bounds. The use of \cc{bndsrch} enables us to accomplish practical sub-object safety at a 7\% average slowdown (see \fref{s:eval}).

With the \cc{bndsrch} implementation, we develop library functions for our sub-object tracing collector, as shown in \fref{fig:utility}. We then insert \cc{wyfy\_mutate()} when a sub-object is indexed (see line 8 in \fref{fig:dps}d) and \cc{wyfy\_bndclr()} when deallocating an object (see line 7 in \fref{fig:dps}a and line 5 in \fref{fig:dps}b). With a sub-object tracing table (STT), declared in the global scope, we maintain the mapping information between objects and their sub-objects at runtime. The STT is indexed by a PAC, and each entry holds the head of a linked list for the corresponding PAC. When a sub-object we want to protect is accessed, we first search for the sub-object's bounds in the HBT using a \cc{bndsrch} instruction. If the \cc{bndsrch} returns 1, we just return the signed sub-object pointer. If not, we create a new node, log the connection between them, and add it to the linked list in the STT entry indexed by the parent object's PAC. Then, we store its bounds in the HBT and return the signed pointer.
Note that an inner structure in a nested structure, i.e., a sub-object of the outer structure, can be an object of its own sub-objects. The mapping of such a nested structure can be implicitly maintained in the STT, as shown in \fref{fig:stt}. }

\ignore{
\subsection{Dynamic Pointer Mutation}
\label{ss:dynamic_pointer_mutation}

While the proposed DPT method becomes a common approach to protecting all memory types, its guarantee is limited to inter-object protection. Pursuing extended coverage up to fine-grained, intra-object protection, one straightforward intuition can be that if we create a new pair of a pointer tag and narrowed capability for a sub-object and associate the pair with the sub-object, we could enforce fine-grained checks. At this point, two questions arise; 1) how to distinguish sub-objects and assign new pointer tags? and 2) how and when to assign and release narrowed bounds for sub-objects?

To answer the first question, we note that in LLVM IR, a \cc{GEP} instruction is used to index a sub-object from an object. This allows us to detect each sub-object indexing by looking for \cc{GEP} instructions. However, the \cc{GEP} instruction calculates the address of the indexed sub-object simply by adding an offset to its base address, and as a result, the sub-object inherits the same pointer tag from the parent object. To differentiate sub-objects and associate unique capability, we choose to re-tag (mutate) pointers returned by \cc{GEP}s to generate different tags.

\begin{figure}[t]
\begin{lstlisting}[language=c++, style=mystyle3, escapechar=!]
// Typedef of struct S is borrowed from Fig 1
void *mallocWrapper(size_t size) {
  void *ptr = malloc(size);
  // Craft custom data structures, etc.
  return ptr;
}
void foo() {
  int *intPtr = mallocWrapper(sizeof(int));
  struct S *strPtr = mallocWrapper(sizeof(struct S));
  strPtr->ovfArr[0] = 'a';
  bar(strPtr); ...
}
void bar(struct S* strPtr) {
  strPtr->ovfArr[1] = 'b'; ...
}
\end{lstlisting}
\caption{Ambiguous type information upon \cc{malloc()}.}
\label{fig:malloc_wrapper}
\end{figure}

\begin{figure}[t]
\begin{lstlisting}[language=c++, style=mystyle3, escapechar=!]
typedef struct stt_node {
  void *ptr;
  struct stt_node *next;
} stt_node;
stt_node *STT[SIZE];    // SIZE is 2^(TAG_SIZE)

void *!\textbf{\_\_dpt\_mutate}!(objPtr, subObjPtr, size, type):
  !\color{codeblue}{tagd}! subObjPtr, type;
  !\color{codeblue}{csrch}! subObjPtr, type;
  if !\color{codeblue}{csrch}! !\textit{returns}! 1:
    return subObjPtr;
  !\color{codeblue}{cstr}! subObjPtr, size; // Store metadata in CMT
  linked_list = STT[getTag(objPtr)];
  !\textit{Create a list with subObjPtr and set as a head list}!;
  return subObjPtr;

void !\textbf{\_\_dpt\_clear}!(ptr, lowBnd, uppBnd):
  !\color{codeblue}{cclr}! ptr;
  !\color{codeblue}{xtag}! ptr;
  linked_list = STT[geTag(ptr)];
  !\color{codered}{foreach}! list in linked_list:
    if lowBnd <= list->ptr <= uppBnd:
      !\textbf{\_\_dpt\_bndclr\_recur}!(list->ptr, lowBnd, uppBnd);
      !\textit{Remove list from linked\_list;}!
  return;
  
void !\textbf{\_\_dpt\_clear}!(objPtr, size):
  lowBnd = !\color{codeblue}{xpacm}!(objPtr);
  uppBnd = lowBnd+size;
  !\textbf{\_\_dpt\_clear\_recur}!(objPtr, lowBnd, uppBnd);
  return;
\end{lstlisting}
\caption{Sub-object tracing collector library functions.}
\label{fig:utility}
\end{figure}

Regarding the second question, one possible approach would be to store the capability metadata of all sub-objects at the time of object allocation and clear all the metadata at the time of deallocation. However, we observe that such a static de-/allocation method could be viable for stack and global objects but not for heap objects. This is because dynamic memory de-/allocation methods often have ambiguous types\footnote{Prior LLVM analysis supports runtime type checking for \cc{malloc()}, but the method incurs a runtime overhead and has limited coverage~\cite{llvmtypegithub}.}, and this makes it hard to determine sub-object's bounds at the time of object allocation. \fref{fig:malloc_wrapper} shows an example of \cc{malloc()} wrapper function usage, which can be found in several SPEC workloads, e.g., \cc{povray\_r/s}. In this example, the wrapper function allocates a memory region with a given size without knowing its type, and allocated objects' types are determined at call sites, i.e., outside the wrapper function.

With this observation, we choose to (dynamically) assign sub-object metadata when sub-objects are indexed using \cc{GEP}s. We call this method \textit{dynamic pointer mutation}. Note that \cc{GEP}s provide sub-objects' types based on the static data-type layout, and this allows calculating the size of indexed sub-objects using the known data type. However, one challenging point is that the same sub-object can be indexed multiple times in different locations (see lines 10 and 14 in \fref{fig:malloc_wrapper}). Hence, not to keep creating duplicates of the same metadata, we need to keep track of metadata allocations for sub-objects and allocate new ones only when needed. Such behaviors remind us of garbage collectors (GCs) that trace the allocation and release of memory using a defined data structure, e.g., a linked list, and reclaim unused memory at runtime. Motivated by the GCs, we design a sub-object tracing collector that enables precise and efficient sub-object metadata management.

\noindent
\textbf{Intra-object protection.} Typical GCs implement a linked list to maintain memory reference information. However, in our evaluation, such an approach is shown to impose an intolerable slowdown (up to $\sim$3x) as typical GCs do. Not to be overwhelmed by the overhead, we propose the use of a new \cc{csrch} instruction (see \fref{ss:isa_extensions}). Having a similar mechanism as the \cc{cstr} and \cc{cclr}, the \cc{csrch} takes the address and size of an indexed sub-object, accesses the location indexed by the mutated pointer's tag in the CMT, and efficiently checks the existence of the metadata. The use of \cc{csrch} enables us to accomplish practical sub-object safety at a ?\% average slowdown (see \fref{s:eval}).

With the \cc{csrch} implementation, we develop library functions for our sub-object tracing collector, as shown in \fref{fig:utility}. We then insert \cc{\_\_dpt\_mutate()} when a sub-object is indexed (see line 8 in \fref{fig:dps}d) and \cc{\_\_xtag\_ptr()} when deallocating an object (see line 7 in \fref{fig:dps}a and line 5 in \fref{fig:dps}b). With a sub-object tracing table (STT), declared in the global scope, we maintain the mapping information between objects and their sub-objects at runtime. The STT is indexed by a pointer tag, and each entry holds the head of a linked list for the corresponding tag. When a sub-object we want to protect is accessed, we first search for the sub-object's metadata in the CMT using a \cc{csrch} instruction. If the \cc{csrch} returns 1, we just return the tagged sub-object pointer. If not, we create a new node, log the connection between them, and add it to the linked list in the corresponding STT entry. Then, we store its metadata in the CMT and return the tagged pointer. Note that an inner structure in a nested structure, i.e., a sub-object of the outer structure, can be an object of its own sub-objects. The mapping of such a nested structure can be implicitly maintained in the STT, as shown in \fref{fig:stt}. 

\begin{figure}[t]
\centering
\includegraphics[width=1\columnwidth]{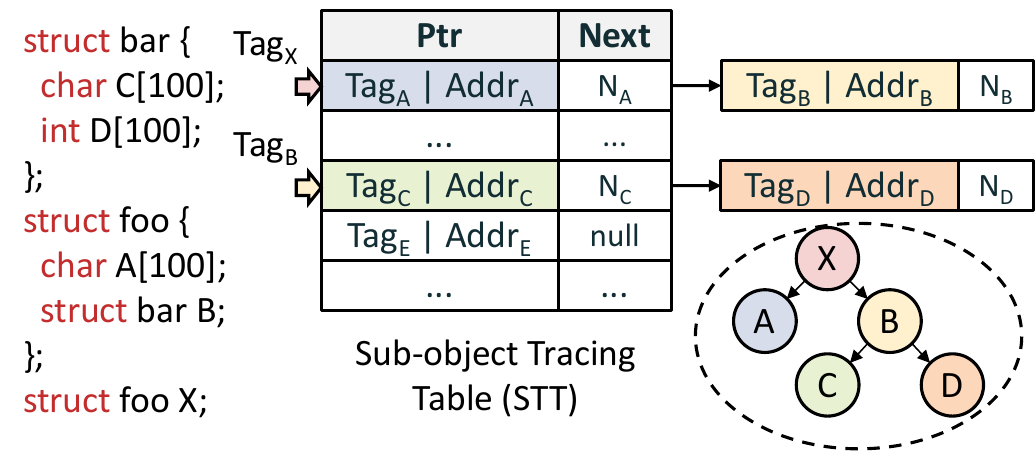}
\caption{Sub-object tracing table (STT).}
\label{fig:stt}
\end{figure}

When an object is deallocated, we erase its mapping information in the STT. By iterating over the linked list corresponding to the object's tag, we look for a list of which pointer address points to the object's memory region. If found, we remove the list from the linked list. At this point, we recall that an inner structure of a nested structure can have its own sub-objects. To clear the traces of all derivative sub-objects, we recursively iterate over the linked list corresponding to the erased pointer's tag. Compared to \cc{\_\_dpt\_mutate()}, \cc{\_\_dpt\_clear()} is called one time for each deallocation, and thus its linked-list iteration overhead is not significant.
}

%% file: Tex/G.Architecture.tex
\section{RV-CURE Architecture}
\label{s:hardware}

For optimal performance, we implement architecture support for DPT. To explore realistic design space, we set the RISC-V BOOM core as our baseline. On top of it, we investigate lightweight hardware extensions, build a synthesizable custom processor, and conduct full-system level evaluations. 

\fref{fig:rv_cure} illustrates the overview of our hardware modifications made on four pipeline stages; \textit{decode}, \textit{execute}, \textit{memory}, \textit{commit} stages. \ballwhite{1} In the decode stage, we detect memory instructions and tag them with additional one bit, namely \cc{needCC}. We use this to indicate that this instruction may perform a potentially unsafe memory access and therefore need to be validated via a capability check. \ballwhite{2} To enable capability checks, in the memory stage, we implement a capability-execution pipeline responsible for capability metadata management and checks. \ballwhite{3} Next, we extend the re-order buffer (ROB) with one-bit entries holding \cc{needCC} bits. Using the extended bit, we ensure that memory instructions commit only after they are validated and turn out to be safe. \ballwhite{4} Lastly, in the execute stage, we add a pointer-tagging unit for new instructions, i.e., \cc{tagd} and \cc{xtag}. We detail our changes as follows.

\subsection{Hardware-based Capability Enforcement}
\mech{} guarantees memory safety via strict capability enforcement in hardware. To achieve this, in the decode stage, we first distinguish potentially unsafe memory instructions, i.e., load and store, and set their \cc{needCC} bits to \cc{true(1)}. By default, the \cc{needCC} bit is set to \cc{false(0)} for other instructions. The identified instructions are propagated through the CPU pipeline and are required to be validated by capability checks. To further strengthen the security level, we extend the ROB with one-bit entries holding \cc{needCC} bits and append one more condition to the existing commit condition such that instructions can commit only when their \cc{needCC} bits are \cc{false(0)}. To support this, we extend the existing interface from the load-store unit (LSU) to the ROB with signals for clearing \cc{needCC} bits, which are asserted to high upon successful validation.

This new constraint enables a precise exception for capability faults, preventing illegal reads and writes from leaking or corrupting memory contents before capability faults are detected and handled. However, in the case where capability checks take longer than normal executions, this can delay instruction retirement at the ROB, thereby degrading performance. For instance, REST~\cite{REST} reported a 23\% higher slowdown with its precise exception support. Even with such a strict policy, we show that \mech{} sustains high performance in \fref{ss:performance}.

\subsection{Capability-Execution Pipeline}\label{cap_exec_pipe}
Furthermore, we design a capability-execution pipeline that enables efficient capability operations. As shown in \fref{fig:rv_cure}, it consists of shadow load and store queues (SLQ and SSQ, respectively), arbitration logics for capability load and store requests, and comparison logics for capability checks. Being paired with a load queue (LQ) and a store queue (SQ), respectively, the SLQ and SSQ take incoming memory addresses, passed from the memory address calculation unit at the execute stage, and verify their accesses.

In \mech{}, selective capability checks are performed for memory accesses by tagged pointers. Hence, when each entry of either the SLQ or the SSQ takes a memory address, it first checks if the address contains a nonzero pointer tag. Only when a nonzero tag is found, each entry proceeds with a capability-check procedure. To verify a memory access, a capability load request is generated to bring in the corresponding metadata from memory. Once the metadata arrive, a capability check is performed, i.e., by checking if the memory address is within the allowed range and by checking if the access is legitimate by the permission bits. Each capability check computes a capability address using the equation (1), where \textit{Base} is the base address of the CMT, \textit{T} is a pointer tag, \textit{N} is the number of ways of the CMT, and \textit{W} is the way to access. \begin{align}
CapAddr = Base + (T << (3 + log_{2}N )) + (W << 3)
\end{align}

Note that the CMT, the storage for capability metadata, has a set-associative structure as mentioned in \fref{ss:ensuring}. Hence, it can accommodate multiple metadata for each tag. Given such a structure, we implement a hardware-based iterative search mechanism, which allows a capability check to iterate over multiple ways without an extra instruction overhead. Hence, its operation is atomic and very cheap.

During an iterative search, if a capability check succeeds, it stops there. Then, it asserts signals for clearing the \cc{needCC} bit to notify the ROB that the corresponding memory instruction is safe to retire. After the instruction commits from the ROB, its corresponding SLQ or SSQ entry is deallocated. If a capability check fails, it calculates a new address with an incremented or decremented way (\textit{W'}) and attempts validation again.

\ignore{
Given the multi-way structure, each capability check computes a capability address using the below equation (1), where \textit{Base} is the base address of the CMT, \textit{T} is a pointer tag, \textit{N} is the number of ways of the CMT, and \textit{W} is the first way to access. \begin{align}
CapAddr = Base + (T << (3 + log_{2}N )) + (W << 3)
\end{align}

When we let \textit{W} the first way to access among the \cc{N} ways of the CMT, a capability address is computed using the below equation, where \textit{Base} is the base address of the CMT, \textit{T} is a pointer tag.\begin{align}
CapAddr = Base + (T << (3 + log_{2}N )) + (W << 3)
\end{align}

For each capability check, 

. Given its set-associative structure, a capability address is computed using the below equation (1), where \textit{Base} is the base address of the CMT, \textit{T} is a pointer tag, \textit{N} is the number of ways of the CMT, and \textit{W} is the way to access. \begin{align}
CapAddr = Base + (T << (3 + log_{2}N )) + (W << 3)
\end{align}

process them based on a finite-state machine (FSM) described below.

\begin{compactitem}[$\bullet$]
  \item \cc{S\_INIT.} Each queue entry starts in this state. If its corresponding memory address arrives, it checks if the address contains a nonzero pointer tag. If found, it transitions to the \cc{S\_READY} state , otherwise to the \cc{S\_DONE} state.
  \item \cc{S\_READY.} An entry in this state generates a capability load request to access its corresponding capability metadata stored in the CMT. Once the request is sent to the data cache (D-cache), it goes to the \cc{S\_WAIT} state.
  \item \cc{S\_WAIT.} An entry in this state waits until a cache response for a capability load arrives. Upon the response arrival, a capability check is performed. If the check passes, it asserts signals for clearing its \cc{needCC} bit sent to the ROB and transitions to the \cc{S\_DONE} state. If not, it goes back to the \cc{S\_READY} to access the next way in the CMT.
  \item \cc{S\_DONE.} In this state, an entry waits until it is deallocated.
\end{compactitem}
}

Besides memory instructions, the SSQ takes and handles new instructions for metadata management, i.e., \cc{cstr} and \cc{cclr}. Those follow the same procedure only except that a \cc{cstr} iterates until empty space is found, and a \cc{cclr} iterates until the expected metadata to clear are found. In addition, an atomic store is additionally performed to the memory location to store or clear metadata. Their operations are similar to a compare-and-swap (CAS) atomic operation.


A capability check may not succeed until iterating over the entire ways of the CMT. This can happen when an illegal memory access is given (therefore, no valid metadata are found). It can also happen when 1) the corresponding set of the CMT is full for \cc{cstr}s or 2) no valid metadata to clear are found for \cc{cclr}s. In these cases, capability faults are generated and handled by an exception handler we introduce in \fref{s:runtime}.

Notably, the hardware complexity we impose on the existing memory pipeline is low. For the capability execution, we only require new capability requests to be scheduled through the LSU scheduler responsible for arbitrating various memory request types. Past the LSU scheduler, capability requests will go through the existing path all the way to the data cache (D-cache) without extra changes. We confirm the negligible impact of our modifications in our RTL synthesis task by being able to synthesize our design at the default clock frequency of \cc{LargeBoomConfig} in FireSim~\cite{firesim}.

However, sharing the execution path can impact performance since it can cause hardware resource contentions between regular memory requests and capability requests. We find three major contention points. First, at the input port of the LSU scheduler, capability requests can conflict with regular memory requests. Second, in the translation lookaside buffer (TLB), capability requests can miss and occupy miss status holding register (MSHR) entries. Given that the TLB cannot accept new requests while its MSHR entries are full, extra TLB misses caused by capability requests can stall the TLB unit until previous misses are resolved. Last, capability requests can miss and occupy MSHR entries in the D-cache, causing stalls on the D-cache until cache misses are resolved.

To minimize performance impact due to such contentions, we take two approaches. First, we assign the lowest priority with capability requests at the LSU scheduler so that regular memory requests are always prioritized. This has an effect of opportunistically utilizing unused cache access bandwidth for capability checks, which minimizes the impact on normal program execution. Second, we introduce small on-chip buffers that help reduce capability requests themselves, explained in the following sections.


\begin{figure}[t]
\centering
\includegraphics[width=0.9\columnwidth]{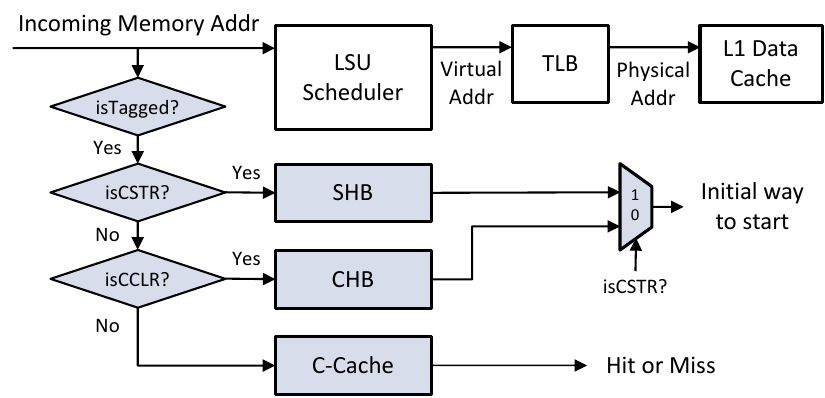}
\caption{Capability cache and store/clear head buffers.}
\label{fig:buffers}
\end{figure}

\subsection{Capability Cache} \label{ss:cache}
DPT assigns a set of metadata regardless of the sub-/object size. This property provides high data locality, and hence even a small storage can be very useful. Based on this, we introduce a capability cache (C-cache) indexed by a lower 8-bit tag, i.e., \cc{tag(7,0)}. The cache consists of a meta array holding a higher 8-bit tag, i.e., \cc{tag(15,8)}, and a data array holding 8-byte metadata, totaling a 2.25KB size. The entries of the C-cache are allocated either when a \cc{cstr} commits its store or when a capability check for a load/store succeeds. When a \cc{cclr} commits, it clears its corresponding entry in the C-cache. As shown in \fref{fig:buffers}, the C-cache is placed in parallel with the LSU scheduler and takes incoming memory addresses. If an incoming address contains a nonzero tag, the C-cache is looked up for metadata satisfying its capability check. If found, i.e., a C-cache hit occurs, the signals for clearing its \cc{needCC} bit are sent to the ROB, and the rest of the check procedure is skipped. 

\subsection{Store/Clear Head Buffers} \label{ss:buffer}
The set associative CMT may require iterative accesses. To optimize its access pattern, we introduce store and clear head buffers (SHB and CHB), as shown in \fref{fig:buffers}. The SHB maintains pointers to empty locations for \cc{cstr}s, and the CHB tracks locations where metadata to clear are stored for \cc{cclr}s. These buffers have a simple array structure indexed by a partial pointer tag, i.e., \cc{tag(7,0)}, totaling a 0.625KB size.

To enable these buffers, we consider the C/C++ programming models. In case of stack objects, last-allocated ones are deallocated first (imagine stack objects belonging to the current stack frame). In contrast, the lifetime of heap objects is not deterministic. Hence, we assume older memory objects will be freed earlier than younger objects. Global objects are never deallocated. Based on this, we set the entry update policy. Below, we assume when a cstr or a cclr has finished its iterative search in the \cc{N}-th way in the \cc{M}-way CMT.

For stack and heap objects, we make a \cc{cstr} update its corresponding SHB entry with \cc{(N+1)\%M} when it commits from the SSQ. When a \cc{cclr} commits, we update its corresponding CHB entry with \cc{(N-1)\%M} for stack objects, and with \cc{(N+1)\%M} for heap objects.
To determine object types, we roughly check if an object address belongs to the upper half of the memory space. If so, it is considered of stack memory type, and otherwise heap memory type (note that stack memory grows from top, and heap memory grows from bottom).

\ignore{
According to the C/C++ programming models, stack objects have a deterministic lifetime; when a callee function is called or returns, its stack objects are allocated or deallocated in a stack frame. Based on this observation, we make a \cc{cstr} update its corresponding SHB entry with \cc{(N+1)\%M} when it commits from the SSQ. When a \cc{cclr} commits, we make it update its corresponding CHB entry with \cc{(N-1)\%M}.

In contrast, the lifetime of heap objects are not deterministic. Hence, we assume that older memory objects will be freed earlier than younger objects. Based on this assumption, we make a \cc{cstr} update its SHB entry with \cc{(N+1)\%M} and a \cc{cclr} update its CHB entry with \cc{(N+1)\%M} when they commit from the SSQ.

make an intuition that if we can maintain pointers pointing to empty locations for \cc{cstr}s and locations where metadata to clear are stored for \cc{cclr}s, their accesses will succeed at their first attempts. With such an intuition, we introduce store and clear head buffers, SHB and CHB in short, shown in \fref{fig:buffers}. These buffers have a simple array structure indexed by a partial pointer tag, i.e., \cc{tag(7,0)}, totaling a 0.625KB size. Each entry of the SHB stores a (likely) empty location for each partial tag, and each entry of the CHB stores a location where the metadata for the next \cc{cclr} are (likely) located. According to the C/C++ programming models, we come up with three possible iteration modes. Since capability metadata for global objects are allocated once and not deallocated during program execution, we focus on stack and heap objects. Below, we assume a situation when a \cc{cstr} or a \cc{cclr} finishes its iterative search in the \cc{N}-th way in the \cc{M}-way CMT.
}

\ignore{
\noindent
\textbf{LAFD (last-allocated, first-deallocated) mode.} Stack objects have a deterministic lifetime; when a callee function is called or returns, its stack objects are allocated or deallocated in a stack frame. Based on this observation, we make a \cc{cstr} update its corresponding SHB entry with \cc{(N+1)\%M} when it commits from the SSQ. When a \cc{cclr} commits, we make it update its corresponding CHB entry with \cc{(N-1)\%M}.

\noindent
\textbf{FAFD (first-allocated, first-deallocated) mode.} In contrast, the lifetime of heap objects are not deterministic. Hence, we assume that older memory objects will be freed earlier than younger objects. Based on this assumption, we make a \cc{cstr} update its SHB entry with \cc{(N+1)\%M} and a \cc{cclr} update its CHB entry with \cc{(N+1)\%M} when they commit from the SSQ.
}
\ignore{
\noindent
\textbf{Adaptive mode.} As a hybrid approach, an adaptive mode takes advantages of both modes. We note that stack memory space grows downward, and heap memory space grows upward in memory space. Based on this, if a memory address of a \cc{cstr} or \cc{cclr} belongs to the upper half of the memory space, the LAFD mode suitable for stack objects is used. Otherwise, the FAFD mode suitable for heap objects is used.

To evaluate the effectiveness of such modes, we run SPEC 2017 C/C++ workloads using reference inputs and measure the average number of iterations for each mode in Column 4-6 in \fref{tab:hit_iter}. Interestingly, the results allow us to categorize the evaluated workloads into three types. First, workloads that do not incur frequent memory allocations, e.g., \cc{lbm}, exhibit near-one average iteration numbers for both modes since there are almost zero hash collisions. Second, workloads that invoke frequent dynamic memory allocations, e.g., \cc{perlbench\_r}, \cc{mcf\_r}, and \cc{xalancbmk\_r}, have lower iteration numbers with the LAFD mode. Third, workloads that invoke frequent function calls leading to frequent stack object de-/allocations, e.g., \cc{nab\_r}, show lower iteration numbers with the FAFD mode. Notably, the adaptive mode shows the lowest iteration numbers in all workloads. Thus, we choose the adaptive mode as our default mode. For comparisons, we also report the results of the baseline mode (\cc{Base}) where an iteration always starts from the first way and the AOS's iteration mode (\cc{BWB}) in Columns 2 and 3 in \fref{tab:hit_iter},. The baseline mode presents 4.37 iterative accesses on average, justifying the need for optimization. The BWB mode shows 3.75 iterative accesses on average, worse than the results of all the LAFD, FAFD, and adaptive modes.
}

\subsection{Handling Tag Dependencies}
Being consistent with the nature of out-of-order (OOO) execution of the BOOM core, we allow OOO execution of capability checks. However, this mandates correct ordering between capability checks and stores sharing the same tag,
i.e., tag dependencies matter. To handle memory dependencies for regular loads and stores, the BOOM core implements costly hardware logic that compares the memory address of a fired load with all stores older than the load and kills the load if any older store with the same address is found. Fence instructions can also be used to prevent the execution of following instructions at the cost of non-trivial pipeline stalls. As a balanced way, we take an approach of stalling capability checks of instructions until all capability store instructions older than themselves, regardless of their tag values, commit from the SSQ. Despite such a conservative approach, the number of capability stores is much less than the number of memory instructions, and both hardware and performance overheads are shown to be trivial.

%% file: Tex/H.Runtime.tex
\begin{table}[t]
    \centering
    \caption{New control and status registers in \mech{}.}
    \begin{scriptsize}
    \begin{tabular}{@{}ll@{}}
    
    \toprule
    \textbf{Name}                        & \textbf{Description}                           \\
    \midrule
    enableDPT                   & Switch to turn on capability enforcement             \\
    baseAddrCMT                 & Base address of a CMT               \\
    numWaysCMT                  & Number of ways of a CMT             \\
    \bottomrule
    \end{tabular}
\label{tab:csrs}
\end{scriptsize}
\end{table}

\section{Runtime Support}\label{s:runtime}
To operate \mech{} in a real system, we add runtime support to the Linux kernel compatible with RISC-V, consisting of an exception handler for capability faults, new control and status registers (CSRs) for hardware configurations, and new process fields for process management.

\noindent
\textbf{Capability fault handler.} To handle capability faults, we define a new exception handler that has a similar mechanism with a page fault handler. Upon a capability fault exception, the exception handler reads the \cc{cause} CSR and determines which instruction type triggered the exception. If a \cc{cstr} instruction turns out to fault, the kernel resizes the CMT by increasing the number of ways and rearranges capability metadata in the CMT. If a \cc{cclr} instruction faults, this may indicate a temporal safety error, such as double free. If a load or store faults, this indicates a capability-check failure. In the last two cases, errors are printed to users, and the execution resumes.

\noindent
\textbf{New CSRs.} To configure the CPU pipeline for \mech{}, we define new CSRs in the standard RISC-V ISA~\cite{RISC-V-ISA}, as shown in \fref{tab:csrs}. The \cc{enableDPT} CSR is used as a switch to turn on the \mech{} mode. The \cc{baseAddrCMT} and \cc{numWaysCMT} CSRs hold the base address and the number of ways of a CMT, respectively. To interface with such CSRs, we add a custom system call, \cc{\_\_dpt\_set()}, that is inserted into a program entry at instrumentation time. In addition to those, we add extra CSRs for debugging and gathering runtime statistics. If \cc{enableDPT} is set to false, all mechanisms of \mech{} are disabled and thus do not impact the normal CPU execution.


\noindent
\textbf{New process fields.} Since all configurations we define above are process-specific, the kernel is required to keep track of each user process' information. To do so, we add new fields to the process structure, i.e., \cc{task\_struct}, that are initialized upon process creation and are set by our custom system call. Upon a context switch, the kernel saves the current process' configurations into the new process fields if the current process is enabled with \mech{} and configures the CPU pipeline with the next process' configurations if the next process needs to be enabled for \mech{}.


%% file: Tex/I.Pruning.tex
\ignore{
\section{DPT with Static Taint Analysis}
\label{ss:taint}

\begin{table}[t]
\scriptsize
\centering
\caption{Static taint analysis results on SPEC CPU 2017.}
\label{tab:taint_results}
\begin{tabular}{@{}lrrrrrr@{}}
    \toprule
    \multirow{2}{*}{\textbf{Name}}
        & \multicolumn{2}{c}{\textbf{Stack Objects}}
        & \multicolumn{2}{c}{\textbf{Heap Objects}} 
        & \multicolumn{2}{c}{\textbf{Global Objects}} \\
        
    \cmidrule(lr){2-3}
    \cmidrule(lr){4-5}
    \cmidrule(lr){6-7}
        & \textbf{\# Taint} & \textbf{\# Total}
        & \textbf{\# Taint} & \textbf{\# Total}
        & \textbf{\# Taint} & \textbf{\# Total} \\
    \midrule
    \textbf{perlbench\_r}   & 184       & 220       & 53        & 67     & 52     & 214     \\
    \textbf{gcc\_r}         & 1195      & 1660      & 80        & 90     & 374     & 894     \\
    \textbf{mcf\_r}         & 1         & 4         & 0         & 18    & 1     & 1    \\
    \textbf{namd\_r}        & 17        & 27        & 88        & 92   & 0     & 0     \\
    \textbf{parest\_r}      & 7135      & 8762      & 680       & 974   & 37     & 2215     \\
    \textbf{povray\_r}      & 949       & 993       & 52        & 69    & 19    & 158   \\
    \textbf{lbm\_r}         & 3         & 9         & 0         & 1    & 0    & 2    \\
    \textbf{omnetpp\_r}     & 1166      & 1515      & 243       & 834    & 146    & 819   \\
    \textbf{xalancbmk\_r}   & 828       & 920       & 16        & 31    & 208     & 2620     \\
    \textbf{x264\_r}        & 110       & 205       & 17        & 21     & 55     & 234     \\
    \textbf{blender\_r}     & 8026      & 8767      & 63        & 96    & 129     & 13659     \\
    \textbf{deepsjeng\_r}   & 23        & 36        & 1         & 1     & 59     & 77     \\
    \textbf{imagick\_r}     & 365       & 430       & 1         & 4    & 11     & 131     \\
    \textbf{leela\_r}       & 124       & 188       & 18        & 28    & 6     & 10     \\
    \textbf{nab\_r}         & 105       & 145       & 27        & 27    & 5     & 25     \\
    \textbf{xz\_r}          & 39        & 54        & 9         & 9    & 11     & 36     \\
    \midrule
    \textbf{Avg. Ratio}
        & \multicolumn{2}{c}{70.34\%}
        & \multicolumn{2}{c}{63.43\%}
        & \multicolumn{2}{c}{26.60\%} \\
    \bottomrule
\end{tabular}
\end{table}

Extending security coverage causes inevitable side-effects: the memory overhead by object-granular metadata and the execution overhead by extra security checks. To mitigate such overheads, we support static taint analysis that helps identify security-critical objects and remove unnecessary checks. Our LLVM optimizer passes consist of points-to analysis, taint analysis, and selective DPT passes.

\ignore{
\subsection{Terminology}
We define the terms we use to describe our method.

\begin{compactitem}[$\bullet$]
  \item A \textit{call graph (CG)} records the function call relationship in a program. A \textit{control-flow graph} (CFG) represents the control flow between \textit{basic blocks} (BBs), and each BB contains a sequence of instructions without any branch except for the exit. For program analysis, we obtain a CG using the existing \cc{CallGraphWrapper} pass and topologically sort each function's CFG using the Tarjan's algorithm~\cite{Tarjan72}. During our analysis, we iterate over instructions by traversing all BBs in functions.
  
  \item In LLVM, the \textit{value} class is the base class of all values, including instruction and function. LLVM constructs the \textit{def-use} and \textit{use-def} chains, where each value maintains a list of values using the particular value. For example, if \cc{value B} uses \cc{value A} as an operand, it becomes a \textit{user} of \cc{value A}. By iterating over the def-use chain, e.g., \cc{A->users()}, we keep track of the possible data flows of values.
\end{compactitem}
}

\ignore{
\begin{figure}[t]
\begin{lstlisting}[language=c++, style=mystyle3, escapechar=!]
visitSet = {}; // Init
void getPtrAlias(V): // V is a pointer value
  if V is found in visitSet: return;
  else: visitSet.insert(V);
    
  !\color{codered}{foreach}! U in V->users(): // U is a user of V
    if U is !\color{codeblue}{GEPInst}!:
      if a succ. node with same indices found:
        !\textit{Add U to alias set of succ. node}!;
      else:
        !\textit{Create new succ. node and add U to the node}!;
      getPtrAlias(U);
    !\color{codered}{elif}! U is !\color{codeblue}{BitCastInst}!:
      if a succ. node with same type found:
        !\textit{Add U to alias set of succ. node}!;
      else:
        !\textit{Create new succ. node and add U to the node}!;
      getPtrAlias(U);
    !\color{codered}{elif}! U is !\color{codeblue}{StoreInst}! or !\color{codeblue}{LoadInst}!:
      !\textit{Add valOp to alias set of ptrOp}!!\color{black}{'}!!\textit{s mem. node}!;
    !\color{codered}{elif}! U is !\color{codeblue}{CallInst}!:
      !\textit{Add argval to alias set}!;
      getPtrAlias(argval);
    !\color{codered}{elif}! U is !\color{codeblue}{ReturnInst}!:
      !\textit{Add U to alias set}!;
      !\color{codered}{foreach}! retval in call sites:
        !\textit{Add retval to alias set}!;
        getPtrAlias(retval);
  return

void pointsToAnalysis(M): // M is a module
  valueSet = {}; // Init
  !\color{codered}{foreach}! Inst in BBs,Funcs,CG:
    if Inst is !\color{codeblue}{AllocaInst}!:
      valueSet.insert(Inst);
    !\color{codered}{elif}! Inst is !\color{codeblue}{CallInst}!:
      if calledFunc is Dynamic alloc. func:
        valueSet.insert(Inst);
  !\color{codered}{foreach}! GV in M.getGlobalList(): // Global variable
    valueSet.insert(GV);
  !\color{codered}{foreach}! V in valueSet:
    !\textit{Create root node with value}!;
    getPtrAlias(V):
  return;
\end{lstlisting}
\caption{Algorithm of \mname{} points-to analysis.}
\label{fig:points_to}
\end{figure}
}

\noindent
\textbf{Points-to analysis.}
This analysis determines the set of pointers (aliases) that can ever point to the same object in memory. By traversing all basic blocks in a program, we first find the pointer values created by memory allocation methods, e.g., \cc{\%stackObj}, \cc{\%heapPtr}, and \cc{\%call} in \fref{fig:dps}c. Using \cc{getGlobalList()} in a given module, we also obtain a list of global variables, e.g., \cc{@glbObj} in \fref{fig:dps}c. We then create a root node for each value and continue to construct a \textit{points-to graph}, where each node contains a set of aliasing values with the same type. Starting from each root node, we recursively iterate over the users of each value. During the iteration, we create a successor node when a pointer creates a new value or is cast to a different type, e.g., \cc{GetElementPtr}, \cc{BitCast}, \cc{Load}, \cc{Call}, \cc{Return} instructions.

\ignore{
Specifically, we focus on the instruction types below. 

\begin{compactitem}[$\bullet$]
    \item \textit{GetElementPtr} (\cc{\%result = GEP(type,\%ptrOp,indices)}) calculates the address of a sub-object within an object. To be field-sensitive, we maintain the indices for each node that includes sub-objects. When a new sub-object value is indexed, we look for a successor node with the same indices. If found, we put the value in that node's alias set. Otherwise, we create a new successor.
    
    \item \textit{BitCast} (\cc{\%result = bitcast type \%op to type2}) converts the operand's type to a different type (\cc{type2}) and creates a new value. We search for a successor node with the same type. If found, we add the new pointer to the node's alias set. If not, we create a new successor.
    
    \item \textit{Store} (\cc{store \%valOp, \%ptrOp}) writes a value to the memory address pointed to by a pointer operand. Note that pointers can be propagated through memory. Conservatively, we consider all values, stored in or loaded from any alias of a pointer operand, as aliased. To do so, we form a \textit{memory edge} between two nodes containing either a value operand or a pointer operand, i.e., \cc{node A} \cc{->} \cc{node B} when an alias in \cc{node A} is stored in or loaded from the address pointed to by any alias in \cc{node B}. Here, we call \cc{node A} a \textit{memory node} of \cc{node B}. Hence, when a value is used as a value operand of a store instruction, we add the value to the alias set of the pointer operand's memory node.
    
    \item \textit{Load} (\cc{\%valOp = load type, \%ptrOp}) reads a value from the memory address pointed to by a pointer operand. As we do for \textit{Store}, we add the value operand to the alias set of the pointer operand's memory node.
    
    \item \textit{Call} (\cc{\%call = call @name(args)}) indicates a function call that makes a control-flow change. If a pointer value is used as a function call's operand, we obtain its corresponding argument by looking into the function definition and add it to the alias set of the pointer value's node.
    
    \item \textit{Return} (\cc{ret \{\%retval}\}) returns to a caller function with an optional return value. If a pointer value is returned, we search for all call sites and add the return values of the callee functions to the alias set of the pointer value's node.
\end{compactitem}
}

\ignore{
\begin{figure}[t]
\begin{lstlisting}[language=c++, style=mystyle3, escapechar=!]
PG = getPointsToGraph(); // From points-to analysis
visitSet = {}; // Init

void handleNode(node):
  if node is found in visitSet: return;
  else: visitSet.insert(node);
  
  node->setTainted(true); // Mark as tainted
  !\color{codered}{foreach}! alias in node->aliasSet:
    !\color{codered}{foreach}! U in alias->users():
      if U is LoadInst:
        doTaintPropagation(U);
  return;

void doTaintPropagation(V):
  !\color{codered}{foreach}! U in V->users(): // U is a user of V
    if U is !\color{codeblue}{StoreInst}!: // store %valOp, %ptrOp
      if valOp == V:
        handleNode(PG->findNode(ptrOp));
    !\color{codered}{elif}! U is !\color{codeblue}{LoadInst}!:
      continue;
    !\color{codered}{elif}! U is !\color{codeblue}{GEPInst}!:
      handleNode(PG->findNode(U));
    !\color{codered}{elif}! U is !\color{codeblue}{CallInst}!:
      !\textit{Get argval from function definition}!;
      doTaintPropagation(argval);
    !\color{codered}{elif}! U is !\color{codeblue}{ReturnInst}!:
      !\color{codered}{foreach}! retval in call sites:
        doTaintPropagation(retval);
    else:
      doTaintPropagation(U);
  return;
  
void taintAnalysis(M):
  !\color{codered}{foreach}! Inst in BBs,Funcs,CG:
    if Inst is !\color{codeblue}{CallInst}!:
      if calledFunc is Extern. I/O func:
        handleNode(PG->findNode(Inst));
  return;
\end{lstlisting}
\caption{Algorithm of \mname{} taint analysis.}
\label{fig:taint}
\end{figure}
}

\noindent
\textbf{Taint analysis.}
In this analysis, we detect taint sources and propagate taint information throughout a program's data flow. We define taint sources as the destination operands of library functions, e.g., \cc{arr} in \cc{scanf("\%s",arr)}, that receive potentially malicious data, such as file I/O, user inputs, and network. To mark values as tainted, we set a propagation policy such that 1) \textit{any pointers that may hold data coming from tainted pointers} are tainted and 2) \textit{all aliases of a tainted pointer} are also tainted. With such a policy, as shown in \fref{tab:taint_results}, we identify 70.34\%, 63.43\%, and 26.6\% of stack, heap, and global objects in the SPEC 2017 C/C++ workloads. We discuss performance and security implications of the results in \fref{ss:performance} and \fref{ss:use_of_taint}.

\ignore{
We iterate over instructions to search for library function calls that receive external inputs, such as \cc{scanf()}. If found, we find a node to which the function's taint source value belongs. We then determine the node as tainted, iterate over the users of each alias in the node's alias set, and look for load instructions. Note that the loads found should use the taint source value or its tainted alias as a pointer operand. Hence, we proceed with a recursive taint propagation call that looks for values taking data loaded from a tainted pointer and marks them as tainted.

To extend the coverage, we further take into account several cases where static analysis could have limitations.
}

\begin{figure}[t]
\centering
\includegraphics[width=0.8\columnwidth]{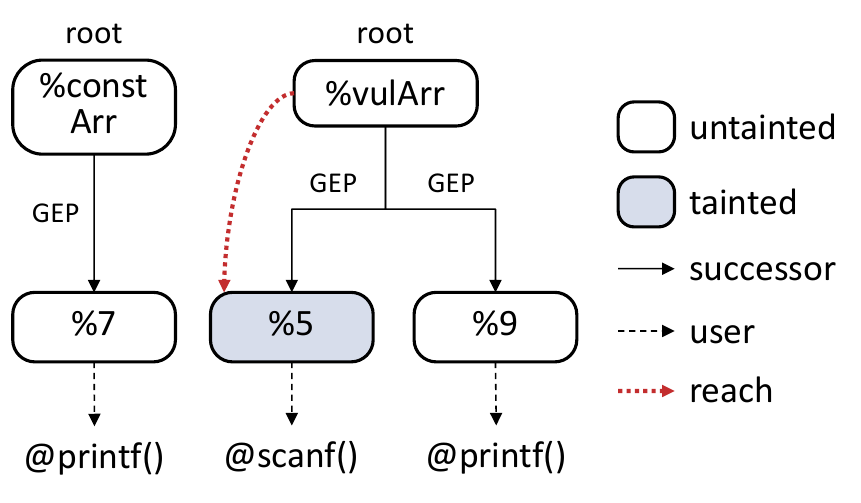}

\centering \small (a) Points-to graph representation during a reachability test.

\begin{lstlisting}[language=c++, style=mystyle3, escapechar=!]
char constArr[] = "Hello World";
char vulArr[10];
!\colorbox{gray!20}{\color{codered}{char} \color{black}{*ptr = vulArr;}\ \ \ \ \ \ \ \ \ \ \ \ \ \ \ \ \ \ \ \ \ \ \ \ \ \ \ \ \ \ \ \ \ \ \ \ \ \ \ \ \ \ \ }!
!\colorbox{gray!20}{\color{codeblue}{tagd} \color{black}{ptr;}\ \ \ \ \ \ \ \ \ \ \ \ \ \ \ \ \ \ \ \ \ \ \ \ \ \ \ \ \ \ \ \ \ \ \ \ \ \ \ \ \ \ \ \ \ \ \ \ \ \ \ \ \ }!
!\colorbox{gray!20}{\color{codeblue}{cstr} \color{black}{ptr, 10;}\ \ \ \ \ \ \ \ \ \ \ \ \ \ \ \ \ \ \ \ \ \ \ \ \ \ \ \ \ \ \ \ \ \ \ \ \ \ \ \ \ \ \ \ \ \ \ \ \ }!
scanf("%s", ptr);
printf("constArr: %s\n", constArr);
printf("vulArr: %s\n", ptr); ... }
\end{lstlisting}

\vspace{0.5cm}
\centering \small (b) Instrumented code after applying the selective DPT pass.
\caption{Example of selective DPT on unsafe code (\fref{lst:example_code}).}
\label{fig:selective_sign}
\end{figure}

\ignore{
\noindent
\textbf{Intrinsic functions.} For better optimization, some library functions can be represented as intrinsic functions. Since their function bodies are not presented in LLVM IR, this hinders our inter-procedural analysis. To increase the coverage, we detect string functions known to be prone to memory bugs, such as \cc{strcpy()} and \cc{memcpy()}, and manually propagate taint from source to destination.

\noindent
\textbf{Function pointers.} The static analysis has difficulties in tracking the control flow change by function pointers since the functions being called are determined at runtime. To mitigate the limitation, we collect functions not included in the CG, which might be called only via function pointers, and perform separate taint analysis, tainting all function arguments.

\noindent
\textbf{Command line arguments.} Since the command line arguments, e.g., \cc{argc}, \cc{argv[]} in C/C++, are supplied by a user, we taint all the arguments of \cc{main()}.
}

\noindent
\textbf{Selective DPT.}
Finally, we determine which values are vulnerable, in terms of memory safety, and thus need to be protected via DPT. To make a decision, we perform a simple taint reachability test. In the points-to graph obtained from the previous analysis, we traverse all successor nodes of each root node and check if any tainted node is found. Note that each root node contains a memory object value returned by an initial memory allocation method. If a tainted node is found, we consider the root node's object value to be vulnerable and therefore apply DPT to the value (see \fref{sss:protection}). Notably, the only decision we make through all the analysis is whether we apply DPT at each object's allocation site or not. Hence, we do not make any intrusive changes to the program code.

\fref{fig:selective_sign} shows an example of code instrumentation using selective DPT. During the taint analysis, we detect the value used as a destination operand of \cc{scanf()} and mark as tainted the node to which the value belongs, as shown in \fref{fig:selective_sign}a. We then figure out the nodes reaching the tainted node. As a result, we recognize \cc{\%vulArr} as a security-critical object and protect it via DPT, as shown in \fref{fig:selective_sign}b.

}

%% file: Tex/V.Methodology.tex
\section{Methodology}

\begin{table}[t]
    \scriptsize
    \centering
    \caption{System configurations of the BOOM core.}
    \begin{tabular}{llll}
      \toprule
      \textbf{Clock}
        & 65 MHz
        & \textbf{L1-I cache}
        & 32KB, 8-way \\
      \textbf{LLC}
        & 4MB
        & \textbf{L1-D cache}
        & 64KB, 16-way \\
      \textbf{DRAM}
        & 16 GB DDR3
        & \textbf{L2 cache}
        & 512KB, 8-way \\
      \midrule
      \multicolumn{2}{l}{\textbf{Front-end}}
        & \multicolumn{2}{l}{8-wide fetch} \\
        & & \multicolumn{2}{l}{16 RAS {\&} 512 BTB entries} \\
        & & \multicolumn{2}{l}{gshare branch predictor} \\
      \cmidrule(lr){3-4}
      \multicolumn{2}{l}{\textbf{Execution}}
        & \multicolumn{2}{l}{3-wide decode/dispatch} \\
        & & \multicolumn{2}{l}{96 ROB entries} \\
        & & \multicolumn{2}{l}{100 int {\&} 96 floating point regs} \\
        \cmidrule(lr){3-4}
        \multicolumn{2}{l}{\textbf{Load-store unit}}
        & \multicolumn{2}{l}{24 LQ and 24 SQ entries} \\
        & & \multicolumn{2}{l}{24 SLQ and 24 SSQ entries} \\
        \cmidrule(lr){3-4}
        \multicolumn{2}{l}{\textbf{Ptr-tagging unit}}
        & \multicolumn{2}{l}{tagd and xtag: 1-cycle latency} \\
        \bottomrule
    \end{tabular}
    \label{tab:params}
\end{table}

We prototype \mech{} on top of the RISC-V BOOM core~\cite{BOOM} and operate using FireSim~\cite{firesim}, an open-source FPGA-accelerated hardware platform for full-system simulation. We synthesize our custom core at 65MHz, the default target frequency of \cc{LargeBoomConfig}, and launch Amazon EC2 F1 instances running the Linux OS compatible with RISC-V. To create instrumented binaries, we design custom optimizer and backend passes in LLVM-9.0.1~\cite{LLVM}. To add runtime support required by \mech{}, we modify the Linux kernel (\cc{version:5.7-rc3)}. To evaluate performance, we compile and run the SPEC 2017 C/C++ workloads~\cite{SPEC2017}. While our environment setup enables cycle-accurate simulations at a full-system level, we observe that the inherently long execution time of some SPEC 2017 C/C++ workloads makes FPGA simulations infeasible. Note that the achievable clock frequency is at most 65MHz due to the limitation of the FPGA environment, and this can lead to simulation time longer than a few weeks for several workloads. Considering this constraint, we estimate the expected execution time by running vanilla programs on our host machine and choose an appropriate input size for each workload that allows us a reasonable simulation time frame (< 1 week). As a result, we run \cc{mcf\_r}, \cc{x264\_r}, and \cc{xz\_r} using reference inputs, run \cc{lbm\_r}, \cc{deepsjeng\_r}, and \cc{nab\_r} using train inputs, and run the others using test inputs. With the system configurations presented in \fref{tab:params}, we evaluate three configurations as follows.

\begin{compactitem}[$\bullet$]
  \item \cc{DPT-H}: Heap protection only. This configuration has security comparable to that of prior work~\cite{AOS,CHEx86,C3}.
  \item \cc{DPT-C}: Coarse-grained protection on all memory types.
  \item \cc{DPT-F}: Fine-grained protection with intra-object protection.
\end{compactitem}

%% file: Tex/W.Evaluation.tex
\section{Evaluation}
\label{s:eval}

\begin{figure}[t]
\centering
\includegraphics[width=\columnwidth]{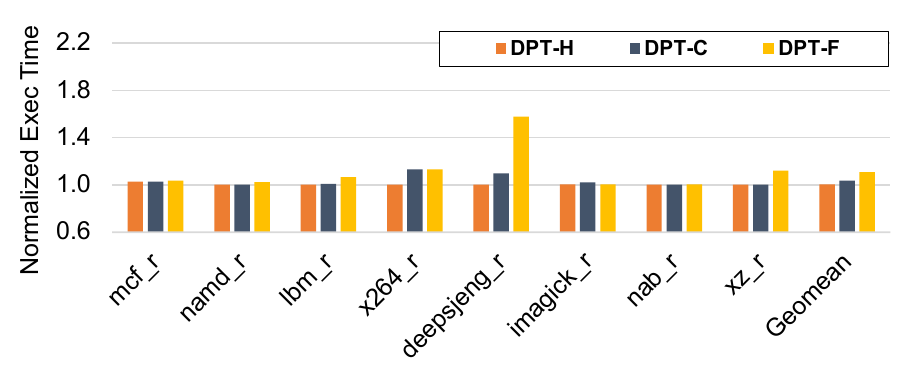}
\caption{Normalized execution time of SPEC CPU2017.}
\label{fig:execution_time}
\end{figure}

\begin{figure}[t]
\centering
\includegraphics[width=\columnwidth]{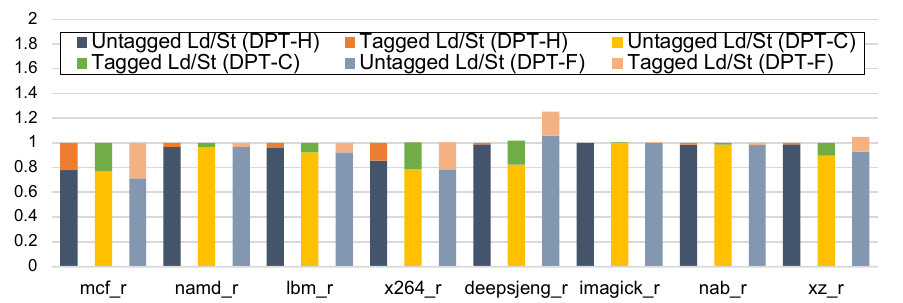}
\caption{Memory instruction overhead and its breakdown.}
\label{fig:load_store}
\end{figure}

\begin{figure}[t]
\centering
\includegraphics[width=\columnwidth]{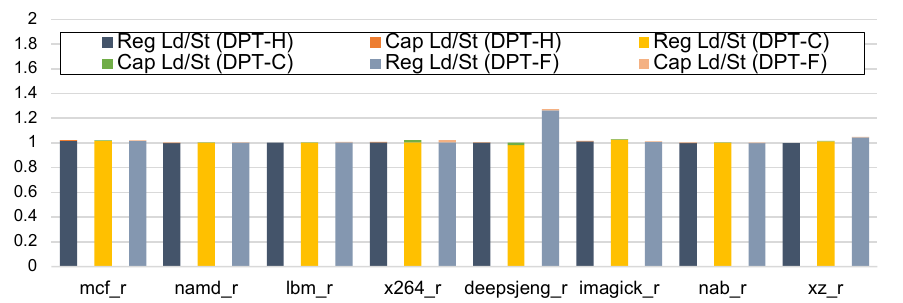}
\caption{Network traffic overhead and its breakdown.}
\label{fig:network_traffic}
\end{figure}

\begin{table}[t]
\scriptsize
\centering
\caption{C-cache hit rates, average iteration numbers of \cc{cstr} and \cc{cclr}, and the number of ways of the CMT measured in DPT-F.}
\label{tab:taint_results}
\begin{tabular}{@{}lcccc@{}}
    \toprule
    \multirow{2}{*}{\textbf{Name}}
        & \multicolumn{1}{c}{\textbf{C-Cache}}
        & \multicolumn{2}{c}{\textbf{\# Avg Iteration}}
        & \multirow{2}{*}{\textbf{\# Max Ways}} \\
    \cmidrule(lr){2-2}
    \cmidrule(lr){3-4}
        & \textbf{Hit Rate (\%)} 
        & \textbf{Base} & \textbf{SHB/CHB} \\
    \midrule
    \textbf{mcf\_r}         & 98.52     & 5.45      & 1.00      & 8                 \\
    \textbf{namd\_r}        & 99.58     & 2.23      & 1.00      & 4                 \\
    \textbf{lbm\_r}         & 99.39     & 1.00      & 1.00      & 4                 \\
    \textbf{x264\_r}        & 98.12     & 1.08      & 1.00      & 12                 \\
    \textbf{deepsjeng\_r}   & 98.65     & 1.05      & 1.00      & 4                \\
    \textbf{imagick\_r}     & 98.79     & 1.00      & 1.00      & 4                 \\
    \textbf{nab\_r}         & 99.39     & 7.43      & 1.00      & 4              \\
    \textbf{xz\_r}          & 99.70     & 1.00      & 1.00      & 8                \\
    \midrule
    \textbf{Average}        & 99.00     & 2.53      & 1.00      & 6                \\
    \bottomrule
\end{tabular}
\label{tab:hit_iter}
\end{table}

\subsection{Performance Analysis}\label{ss:performance}
\fref{fig:execution_time} presents the execution time of three configurations, normalized to the baseline. \cc{DPT-H}, \cc{DPT-C}, and \cc{DPT-F} exhibit 0.5\%, 3.5\%, and 10.8\% average slowdowns, respectively. First, \cc{DPT-H} only enables heap protection. \cc{DPT-C} extends security coverage up to stack and global objects, and \cc{DPT-F} further extends with sub-object safety. As we pursue higher security levels, we observe the average overhead increase. Nevertheless, most applications show negligible overheads except for \cc{deepsjeng\_r}, which we analyze in details as follows.



In general, we identify that the runtime overhead mainly comes from the memory overhead by capability metadata and the execution overhead by capability checks. This is because as security coverage is broadened, the number of memory objects we need to protect increases, and therefore more capability metadata and checks are needed. Given that capability metadata are redundant in terms of normal program execution, such metadata can thrash useful cache lines when they are brought into the D-cache, causing cache pollution. Moreover, the increased number of capability checks for more memory objects can result in increased hardware resource contentions in the memory pipeline.

To further interpret the results, we collect runtime statistics from our FPGA simulations and examine memory execution behaviors. \fref{fig:load_store} shows the memory instruction overhead, compared to the baseline, and the tag ratio of memory instructions. Since we only add tagging and capability instructions, we observe almost zero overheads for \cc{DPT-H} and \cc{DPT-C}, indicating that our compiler instrumentation is not intrusive. Even in \cc{DPT-F}, we only observe noticeable 25\% and 4.8\% overheads for \cc{deepsjeng\_r} and \cc{xz\_r}, respectively. Our code analysis reveals that \cc{deepsjeng\_r} allocates 15 million heap objects of struct type and keeps incurring sub-object indexing during iterations, which requires frequent pointer mutation and card table access operations (see \fref{fig:mutate}). Such an overhead manifests as a 58\% runtime overhead in \cc{deepsjeng\_r}. In \cc{xz\_r}, we also observe the frequent use of various struct types (for its encoder and decoder data structures), which leads to a 12\% runtime overhead. The average tag ratio is measured as 5.93\%, 10.79\%, and 11.18\% in \cc{DPT-H}, \cc{DPT-C}, and \cc{DPT-F}, respectively.


Capability checks can generate extra memory requests and thus cause a network traffic overhead. We estimate that overhead by counting the number of regular memory and capability requests sent to the L1 D-cache. \fref{fig:network_traffic} shows the results normalized to that of the baseline. Notably, we measure negligible overheads for all applications. Only \cc{deepsjeng\_r} and \cc{xz\_r} show the increased number of regular memory requests due to the memory instruction overhead mentioned above.

The results are attributed to the micro-architectural units we introduce. Column 1 in \fref{tab:hit_iter} shows extremely high hit rates of the C-cache, indicating that almost 99\% of checks hit the cache and do not incur extra memory requests. Note that \cc{DPT} assigns only one set of metadata regardless of the sub-/object size, and thus each metadata have high data locality. In addition, Column 2-3 show that the average iteration numbers of \cc{cstr} and \cc{cclr} are optimized to one with the help of SHB and CHB. This indicates most \cc{cstr}s and \cc{cclr}s finish their CMT access at the first attempt, and thus the overhead from an iterative search becomes negligible. The number of ways required to hold all capability metadata is shown in Column 5.

Lastly, we measure the instruction overhead by instructions of new types we insert at compile time. \fref{tab:stats} shows their overhead is negligible on average (< 1\%).

\begin{table}[t]
    \centering
    \caption{Instruction overhead by new instructions measured in DPT-F.}
    \begin{scriptsize}
    \begin{tabular}{@{}lrrrr@{}}
    \toprule
                        & \textbf{\# Total Inst} & \textbf{\# TAGD/XTAG}  & \textbf{\# CSTR/CCLR}  & \textbf{Overhead}     \\
    \midrule
    \textbf{mcf\_r}         & 2723B     & 78B       & 990K      & 2.90\%            \\
    \textbf{namd\_r}        & 1766B     & 22M       & 177K      & <0.1\%            \\
    \textbf{lbm\_r}         & 383B      & 2.6M      & 172       & <0.1\%                 \\
    \textbf{x264\_r}        & 441B      & 506M      & 494M      & 0.22\%                 \\
    \textbf{deepsjeng\_r}   & 1093B     & 19B       & 702M      & 1.88\%            \\
    \textbf{imagick\_r}     & 1.9B      & 114K      & 5625      & <0.1\%                 \\
    \textbf{nab\_r}         & 2484B     & 323M      & 36K       & 0.01\%                 \\
    \textbf{xr\_r}          & 20B       & 38M       & 126M      & 0.80\%               \\
    \midrule
    \textbf{Average}        & --        & --        & --        & 0.72\% \\
    \bottomrule
    \end{tabular}
    \label{tab:stats}
    \end{scriptsize}
\end{table}

\begin{table}[t]
    \centering
    \caption{Hardware overhead estimates.}
    \begin{scriptsize}
    \begin{tabular}{@{}lcc@{}}
    \toprule
                                & \textbf{Area (mm\textsuperscript{2})}  & \textbf{Power Consumption (mW)}        \\
    \midrule
    \textbf{BOOM Core}~\cite{BOOM}       & 6.2812                        & 15.3612           \\
    \textbf{\mech{} Core}                & 6.8266                        & 17.1550           \\
    \bottomrule
    \end{tabular}
\label{tab:hw_overhead}
\end{scriptsize}
\end{table}

\subsection{Hardware Overhead Estimation}
To estimate the hardware overhead imposed by \mech{}, we synthesize our design using Synopsys Design Compiler using 45nm FreePDK libraries~\cite{FreePDK45} at 1GHz clock frequency. To obtain realistic estimates of the SRAM modules, such as L1 D-cache, L1 instruction-cache, and branch predictor (BPD) tables, we black-box such SRAM blocks and use analytical models from OpenRAM~\cite{OpenRAM}. \fref{tab:hw_overhead} shows that \mech{} only incurs 8.6\% area and 11.6\% power overhead.

%% file: Tex/X.Security.tex
\section{Security Analysis}
\label{sec:security}
\ignore{
\subsection{\mname{} Capabilities}
The capabilities of \mname{} are enforced by a security policy that all memory objects protected have corresponding object pointers carrying over encrypted pointer tags, and all memory accesses by the tagged (signed) pointers are validated.

\noindent
\textbf{Spatial safety.} Once an object is allocated, it occupies a valid, restricted memory region in memory. \mname{} ensures that the alive object's bounds are maintained in memory, and any memory access deviating from its allowed range is detected by comparing the memory address with the object's bounds.

\noindent
\textbf{Temporal safety.} After an object is deallocated, any dangling pointers to the region become invalid. \mname{} prevents the use of such dangling pointers, i.e., use-after-free, by removing the object's bounds from the HBT but leaving dangling pointers signed. Afterward, any subsequent use of such pointers will fail in bounds checking because of the lack of the corresponding bounds.

\noindent
\textbf{Double-free.} In glibc 2.26, the new thread local-caching mechanism (\cc{tcache}) exposed a new heap exploit, namely double-free. \mname{} can also prevent double-free attacks. As can be seen in \fref{lst:heap}, we insert the \cc{wyfy\_bndclr()} before \cc{free()} and enforce the pointer's bounds to be cleared using the \cc{bndclr} . When double-free occurs, the execution of the \cc{bndclr} will fail since the corresponding bounds would have been cleared before and will not be found in the CMT.
}

\subsection{Juliet Test Suite}
To evaluate the \mname{} capabilities, we run a subset of the NIST Juliet test suite 1.3 for C/C++~\cite{Juliet-suite}. Out of a total of 118 test categories, we select the categories of the types of errors we expect \mname{} to cover, as enumerated in \fref{tab:juliet}. The selected test cases include all spatial, temporal, and sub-object safety violations. From the selected categories, we exclude several subcategories that require a network connection or depend on randomly generated values since the the BOOM core does not support those. We confirm that \mname{} with \cc{DPT-F} detects all bugs that lie in the categories of our interest.


\subsection{Memory Bugs in SPEC CPU2017}\label{ss:bug-spec}
As we run the SPEC 2017 C/C++ workloads in \mech{}, we detect several bugs summarized in \fref{tab:spec-bug}. In \cc{deepsjeng\_r} and \cc{x264\_r}, we detect array accesses with a negative index computed by a subtraction operator, e.g., \cc{s->sboard[sq-7]} in \cc{deepsjeng\_r}. In \cc{nab\_r}, we also detect intra-object OOBs. It declares an one-byte-sized array in a struct type, i.e., \cc{catspace[1] in struct re\_guts}, and uses it as a variable-sized array by assigning \cc{N} bytes more when allocating the struct type. Prior literature~\cite{IntelMPX} also observed similar violations in SPEC CPU2006. Although this is an intended behavior (thus not a real bug), it violates the C/C++ memory model. Note that all bugs we find can only be detected with fine-grained protection enabled.


\begin{table}[t]
\scriptsize
\centering
\caption{Tested CWE categories in Juliet test suite.}
\label{tab:juliet}
\begin{tabular}{@{}lcccc@{}}
    \toprule
    \multirow{2}{*}{\textbf{Category}}
    & \multirow{2}{*}{\textbf{\# Tests}}
    & \multicolumn{3}{c}{\textbf{\# Tests Detected}} \\
    
    \cmidrule(lr){3-5}
        & \textbf{} 
        & \textbf{DPT-H} & \textbf{DPT-C} & \textbf{DPT-F} \\
    \midrule
    CWE121 Stack Ovf.       & 340   & 0     & 304   & 340     \\ 
    CWE122 Heap Ovf.        & 420   & 384   & 0     & 420    \\ 
    CWE124 Buf. Underwrite  & 496   & 0     & 496   & 496    \\  
    CWE126 Buf. Overread    & 384   & 0     & 384   & 384  \\  
    CWE127 Buf. Underread   & 496   & 0     & 496   & 496    \\  
    CWE415 Double Free    & 480   & 480   & 0     & 480   \\
    CWE416 Use-After-Free   & 394   & 394   & 0     & 394    \\  
    \bottomrule
\end{tabular}
\end{table}

\begin{table}[t]
\scriptsize
\centering
\caption{Memory bugs found in SPEC CPU2017.}
\label{tab:spec-bug}
\begin{tabular}{@{}lll@{}}
    \toprule

    \textbf{Application}       & \textbf{Bug Description} & \textbf{Location} \\
    \midrule
    deepsjeng\_r    & Array access with negative index  & neval.cpp:222          \\ 
    nab\_r          & Variable-sized array is used      & regcomp.c:1321       \\ 
    x264\_r         & Array access with negative index  & slicetype.c:259         \\  
    
    \bottomrule
\end{tabular}
\end{table}

\subsection{Tag Forgeability}
Attackers may try to forge pointer tags to bypass the \mech{}'s mechanism, e.g., via buffer overflow or pointer arithmetic. However, we believe the attack surface exposed is very limited since \mname{} capabilities would prevent spatial and temporal bugs in the first place. However, manipulating the upper bits via pointer arithmetic (e.g., integer overflow) is still possible. To make such attacks less feasible, we take an approach of making tags unpredictable because an attacker will not know which integer value needs to be used without knowing a pointer tag value.

As explained \fref{ss:isa_extensions}, we use a 16-bit pseudo-random value (\cc{LFSR-16}) when tagging stack and heap object pointers. As the LFSR in hardware keeps shifting its value at each CPU clock cycle, unless an attacker accurately computes clock cycle intervals (in ns), knowing the LFSR value would be impossible. In case of global and sub-object pointers, we use a static type identifier as a modifier since we want to obtain the same tag whenever computing it in different places. Thus, those pointer types could be susceptible. To protect those types, enhanced cryptographic algorithms~\cite{QARMA,Gimli} can be used.

\subsection{Use-After-Reallocation Attacks}\label{ss:use-after-reallocation}
Experienced attackers may reuse a dangling pointer to access a memory region that has been reallocated and pointed to by a different pointer, i.e., reusing \cc{ptrA} when \cc{ptrA=malloc();} \cc{->} \cc{free(ptrA);} \cc{->} \cc{ptrB=malloc();}. Using a 16-bit LFSR value for tag generation can prevent such an attack since \cc{ptrB} will be assigned a different tag with that of \cc{ptrA} even though the memory chunk address is the same.




\subsection{False Positives}
Given the limited tag size, hash collisions can cause false positives across data pointers carrying the same tag. However, the available tag size, varying from 11 to 32 bits depending on the virtual address schemes, provides robustness against false positives. For example, to exploit false positives with a 16-bit tag size, adversaries would need to attempt 45425 runs to achieve a 50\% likelihood for a correct guess~\cite{pacitup}. 

\ignore{
\subsection{Pointer Integrity}
\label{ss:pointer_integrity}
Chen et al.~\cite{dop_attack} demonstrate that non-control-data attacks can be realistic attack vectors. Without altering control flow, such data-oriented programming (DOP) attacks exploit data pointers to corrupt application data, including user identity and security-critical configurations. \mname{} effectively protects against such attacks by enforcing an invariant that memory accesses always operate within the allowed range. Besides the DOP attacks, control-flow attacks, such as return-oriented programming~\cite{ROP} and jump-oriented programming~\cite{JOP}, become a promising attack vector. We note that the proposed data-pointer signing scheme is orthogonal to the existing PA-based code-pointer integrity methods~\cite{armpa, pacitup}. Since our design is based on the AArch64 architecture, the Arm PA primitives can be easily enabled for code-pointer signing to protect against control-flow attacks.
}
\subsection{Double-Free} 

In glibc 2.26, the new thread local-caching mechanism (tcache) exposed a new heap exploit, namely double-free. This attack can be prevented in \mech{}. As shown in \fref{fig:dps}b, we insert a \cc{cclr} before \cc{free()} and clear the pointer's capability metadata. When double-free occurs, the \cc{cclr} will fail since the corresponding metadata would have been already cleared and will not be found in the CMT.

\ignore{
\subsection{Use of Static Taint Analysis}\label{ss:use_of_taint}
The security guarantee of typical static analysis techniques is determined by the precision (high coverage with many false positives) and soundness (moderate coverage with all true positives) of the analysis~\cite{drchecker}.
Meanwhile, extensive research in the area has presented many practical solutions. For instance, control-flow integrity (CFI) and data-flow integrity (DFI) solutions have shown their practicality by detecting numerous bugs in real-world programs~\cite{DFI,CFI,PITTYPAT}. By sharing the basic mechanism, our analysis guarantees security comparable to that of prior work. When users want to use \mech{} for testing and debugging, e.g., to catch and fix benign bugs, or prefer better security,
the static taint analysis can be skipped. 
}


%% file: Tex/Y.Discussion.tex
\section{Discussion}
\subsection{Multi-Threading Support}
Although \mech{} is implemented in a single-threaded environment, we believe \mech{} can be employed for multi-threading applications without much changes.  In \mech{}, the CMT is maintained in memory. Hence, all threads belonging to the same virtual memory space can freely share the same storage for capability operations. The per-core C-cache may hold stale metadata in the case where a different core has cleared that metadata. However, under typical memory consistency models, e.g., total store order (TSO), the C-cache can take a cache-line invalidate request sent from a requestor core and invalidate the entries belonging to the cache line region. Moreover, \mech{} is free from the issues derived from non-atomicity of bounds load and check instructions found in Intel MPX~\cite{IntelMPX}. This is because \mech{} performs capability checks atomically in hardware without extra instructions.

\ignore{
\subsection{Compatibility}
From our full-system evaluations, we learn valuable lessons in terms of compatibility issues all pointer- or memory-tagging methods may have. First, we observe that pointer arithmetic using two pointers, e.g., \cc{offset=ptrA-ptrB;} can result in incorrect values when both pointers are tagged and have different tags. As an example, we find this type of operations in programs where custom data structures are defined, and the relative offset between two objects is obtained by subtracting one's address from another's. To avoid this issue, given that LLVM IR converts such pointer arithmetic into two \cc{ptrtoint} instructions and a \cc{sub} instruction, we insert \cc{xtag}s before \cc{ptrtoint}s and remove pointer tags before pointer arithmetic. We presume that this finding may contradict some prior work that claims the advantage of not requiring source code recompilation~\cite{C3,armpa}.

Next, some functions in standard or third-party libraries, such as file I/O functions, may cause compatibility issues when they take tagged pointers as function arguments. We presume that those functions contain pointer arithmetic incompatible with tagged pointers, such as the above-mentioned case. To work around this issue, we whitelist compatible and safe functions and insert \cc{xtag}s to remove tags of pointers passed to library functions not included in our list. In fact, Google observed a similar challenge with Arm memory tagging extensions (MTE) and modified various routines in the Android C library~\cite{GoogleLLC}. We expect such efforts to be made to help mitigate this issue.
}

\ignore{
\subsection{Static vs. Dynamic Taint Analysis}
Typical static taint analysis~\cite{PhASAR,DepTaint,SWAN,ConDySTA} causes an over-tainting problem since it analyzes all possible executions at compile time. In contrast, dynamic taint analysis~\cite{GPU-taint,TaintCheck,TaintBochs,Dytan,TaintDroid,LIFT} can track the exact execution path at the cost of runtime overhead. In this work, while we focus on static taint analysis to avoid extra runtime overheads, we leave the exploration of dynamic analysis techniques as future work.
}

\subsection{Applicability}
Even though \mech{} is prototyped in the RISC-V ecosystem, we believe that our hardware design is generally applicable to other architectures. This is because the BOOM core has primary architecture components common to modern architectures, such as an out-of-order memory-execution pipeline, and our changes are made on top of those.

%% file: Tex/Y.RelatedWork.tex
\ignore{
\begin{table*}[t]
    \centering
    \caption{Comparison between \mname{} and previously proposed memory safety mechanisms.}
    \label{tab:comparison}
    \begin{threeparttable}
    \begin{scriptsize}
    \begin{tabular}{@{}lccccc@{}}
    
    \toprule
    Mechanism & HW/SW-based & Spatial Safety & Temporal Safety & Sub-Object Safety & Compatibility Loss~\tnote{1} \\
    \midrule
    ASan~\cite{asan}& SW     & \cmark    & \cmark    & \xmark    & $-$    \\
    SoftBound~\cite{Softbound}  & SW    & \cmark    & \xmark    & \xmark    & $-$    \\
    Baggy Bounds~\cite{BaggyBounds,BaggyBoundsChecking}    
                                & SW    & \cmark    & \xmark    & \xmark    & $-$    \\
    TaintCheck~\cite{TaintCheck}& SW    & \cmark~\tnote{2} & \cmark~\tnote{2} & \xmark & $-$ \\
    \midrule
    
    REST~\cite{REST}            & HW    & \cmark~\tnote{3}    & \xmark    & \xmark    & $-$    \\ 
    Califorms~\cite{Califorms}  & HW    & \cmark~\tnote{3}    & \cmark~\tnote{3}    & \xmark    & Src, Bin  \\ 
    \midrule
    Intel MPX~\cite{IntelMPX}   & HW    & \cmark    & \xmark    & \cmark    & $-$    \\ 
    HardBound~\cite{Hardbound}  & HW    & \cmark    & \xmark    & \xmark    & $-$    \\ 
    Watchdog~\cite{Watchdog}    & HW    & \cmark    & \cmark    & \xmark    & $-$    \\
    CHERI~\cite{CHERI,Cheri-concentrate,CHERIvoke,Cornucopia}
                                & HW    & \cmark    & \cmark    & \cmark    & Src, Bin    \\ 
    \midrule
    Arm MTE~\cite{armpa}        & HW    & \cmark    & \cmark    & \cmark    & $-$    \\ 
    SPARC ADI~\cite{SPARC_ADI}  & HW    & \cmark~\tnote{4}    & \cmark~\tnote{4}    & \xmark    & $-$    \\
    \midrule
    AOS~\cite{AOS}              & HW    & \cmark~\tnote{4}  & \cmark~\tnote{4}  & \xmark    & $-$ \\
    C\textsuperscript{3}~\cite{C3}  & HW    & \cmark~\tnote{4}  & \cmark~\tnote{4}  & \xmark    & $-$ \\
    No-FAT~\cite{No-FAT}          & HW    & \cmark~\tnote  & \cmark~\tnote  & \cmark~\tnote    & Src, Bin \\
    In-Fat Pointer~\cite{in-fat-pointer} & HW   & \cmark   & \xmark    & \cmark    & $-$    \\
    \midrule
    \mname{}    & HW       & \cmark            & \cmark        & \cmark
                & $-$    \\
    \bottomrule
    \end{tabular}
    \end{scriptsize}

    \begin{tablenotes}
    \begin{footnotesize}
    \item[1] Some prior work requires source modifications or recompiling the libraries, losing source compatibility (Src) or binary compatibility (Bin).
    \item[2] Taint analysis-based security checking is applied only to selected attack targets, such as a jump address and format string.
    \item[3] Canary-based protection is applied.
    \item[4] Protection only applies to heap memory regions.
    
    \end{footnotesize}
    \end{tablenotes}    
    
\end{threeparttable}
\label{tab:related_work}
\end{table*}
}

\ignore{
\begin{table}[t]
    \centering
    \caption{Comparison between \mname{} and previous defense mechanisms for memory safety.}
    \label{tab:comparison}
    \begin{threeparttable}
    \begin{footnotesize}
    \begin{tabular}{lcccc}
    
    \toprule
    Mechanism   & Spatial & Temporal & Sub-Obj. & Comp. \\
                & Safety & Safety & Safety & Loss~\tnote{1} \\
              
    \midrule
    REST~\cite{REST}            & \cmark    & \xmark    & \xmark    & $-$    \\ 
    Califorms~\cite{Califorms}  & \cmark    & \cmark    & \xmark    & Src, Bin  \\ 
    \midrule
    Intel MPX~\cite{IntelMPX}   & \cmark    & \xmark    & \cmark    & $-$    \\ 
    HardBound~\cite{Hardbound}  & \cmark    & \xmark    & \xmark    & $-$    \\ 
    Watchdog~\cite{Watchdog}    & \cmark    & \cmark    & \xmark    & $-$    \\
    CHERI~\cite{Cornucopia,Cheri-concentrate,CHERI,CHERIvoke}
                                & \cmark    & \cmark    & \cmark    & Src, Bin    \\ 
    \midrule
    Arm MTE~\cite{armpa}        & \cmark    & \cmark    & \cmark    & $-$    \\ 
    SPARC ADI~\cite{SPARC_ADI}  & \cmark~\tnote{2}    & \cmark~\tnote{2}    & \xmark    & $-$    \\
    \midrule
    AOS~\cite{AOS}              & \cmark~\tnote{2}  & \cmark~\tnote{2}  & \xmark    & $-$ \\
    C\textsuperscript{3}~\cite{C3}  & \cmark~\tnote{2}  & \cmark~\tnote{2}  & \xmark    & $-$ \\
    No-FAT~\cite{No-FAT}          & \cmark~\tnote  & \cmark~\tnote  & \cmark~\tnote    & Src, Bin \\
    In-Fat Pointer~\cite{in-fat-pointer} & \cmark   & \xmark    & \cmark    & $-$    \\
    \midrule
    \mname{}    & \cmark            & \cmark        & \cmark
                & $-$    \\
    \bottomrule
    \end{tabular}
    \end{footnotesize}

    \begin{tablenotes}
    \begin{footnotesize}
    \item[1] Some prior work requires source modifications or recompiling the libraries, losing source compatibility (Src) or binary compatibility (Bin).
    \item[2] Protection only applies to heap memory regions.
    
    \end{footnotesize}
    \end{tablenotes}    
    
\end{threeparttable}
\label{tab:related_work}
\end{table}
}

\section{Related Work}
\label{s:related_work}
\ignore{
\noindent
\textbf{Software solutions.}
While software approaches~\cite{asan,Softbound,BaggyBounds,BaggyBoundsChecking,TaintCheck,EffectiveSan} typically guarantee a variety of security policies, their versatility comes with significant overhead. For example, AddressSanitizer (ASan)~\cite{asan} stores valid bits for live objects in reserved space (shadow memory) for every 8 bytes and validates memory accesses by checking the valid bits. Combined with a quarantine zone design for preventing temporal reuse of freed regions, ASan presents a 73\% slowdown. 
}

\ignore{
\noindent
\textbf{Pointer authentication.} 
As a commercial technology, Arm PA~\cite{armpa} aims to prevent illegal pointer modification. Using QARMA~\cite{QARMA}, Arm PA places a PAC in a pointer and authenticates the PAC before use to verify its integrity. Using the same primitives, PARTS~\cite{pacitup} further extends the security capabilities by proposing PA-based data-pointer integrity.
ZeRØ~\cite{zero21} implements new memory instructions and a metadata encoding scheme to protect code and data pointers. Despite the low cost of PA-based methods, their security coverage is limited since neither spatial nor temporal safety is fully ensured. 
}

\noindent
\textbf{Trip-wire.}
Trip-wire methods place secret bytes around objects to protect. The secret bytes are not supposed to be accessed by unprivileged normal operations, and any access to the secret bytes is prohibited. REST~\cite{REST} embeds a randomized token near benign buffers and has a comparator in the cache hierarchy to detect the token being accessed. Califorms~\cite{Califorms} leverages unused padding bytes and places security bytes at the byte granularity, minimizing a memory overhead. Trip-wire solutions have limited security coverage despite low-cost protection because they cannot detect out-of-bounds accesses jumping over secret bytes.


\noindent
\textbf{Fat pointer.}
\label{ssc:fat-pointer}
This class extends the pointer size to hold security metadata. The CHERI capability model~\cite{CHERI, Cheri-concentrate, CHERIvoke, Cornucopia} implements a capability co-processor with a set of capability registers (256 bits each at worst case) holding bounds and permission bits. Object accesses are forced to check their bounds and control access using the capability metadata. However, the design requires changes on most system stacks, including the compiler, the language runtime, and the OS, and ends up losing both source and binary compatibility. Similarly, Hardbound~\cite{Hardbound} and Watchdog~\cite{Watchdog} implement register extensions to maintain per-pointer metadata. However, the overhead of explicit memory checking and metadata propagation results in a significant slowdown. Intel Memory Protection Extensions (MPX)~\cite{IntelMPX} implements two-level address translation to access bounds. However, its complicated addressing based on a multi-level bounds table incurs a non-trivial overhead.


\noindent
\textbf{Memory tagging.}
SPARC Application Data Integrity (ADI)~\cite{SPARC_ADI} and Arm Memory Tagging Extension (MTE)~\cite{armpa} are the representative commercial products that implement memory tagging. In those designs, memory allocation creates a tag and colors the corresponding memory region with the tag. When loads and stores access the colored region, the tag present in a pointer is compared with the tag stored in memory. However, the limited tag size, e.g., 4 bits in Arm MTE, causes a high false-positive rate and allows an attacker to brute-force the tag bits via a sufficient number of attempts. 

In addition, CHEx86~\cite{CHEx86} proposes a micro-code level instrumentation. With a set of registers holding the entry and exit points of heap management functions, CHEx86 detects heap de-/allocation and controls memory capabilities. Despite its novel scheme, its protection is limited to heap memory. ZeRØ~\cite{zero21} proposes a novel pointer integrity mechanism. By introducing unique memory instructions, ZeRØ prevents accesses to security-critical pointers, e.g., return address, code/data pointers, using regular memory instructions. However, it only guarantees the integrity of pointers and does not entirely prevent spatial and temporal errors, leaving the possibility of data-oriented programming (DOP) attacks~\cite{dop_attack}.

\ignore{
Lack of stack object protection. Its low-level instrumentation depends on a limited set of new model-specific registers (MSRs), which holds the instruction addresses of the entry/exit points of heap de-/allocation functions. Only when the registered addresses are fetched, the capability information can be generated or freed. With this approach, it will be challenging to support stack object protection since instruction addresses for each stack object allocation cannot be specified without appropriate compiler support. Note that the entire stack memory space for a given function is assigned by decrementing the stack pointer by a proper offset, and the bounds for each stack object is not expressed at the ISA level.
Lack of sub-object protection. Similarly, the nature of its low-level instrumentation does not capture the language-level semantics, such as types and sub-object indexing, and therefore providing sub-object protection will be very challenging.
}

\ignore{
\noindent
\textbf{Pointer tagging.}
\label{ss:pointer-tagging}
C\textsuperscript{3}~\cite{C3} proposes a state-less (requiring no extra metadata) mechanism that entangles data encryption with encoded memory addresses.
Using a binning allocator, No-FAT~\cite{No-FAT} exposes memory allocation size to hardware and performs hardware-based bounds checking. To enforce temporal safety, No-FAT stores a random tag in a pointer and verifies the tag upon a memory access. In-Fat Pointer~\cite{in-fat-pointer} diversifies object metadata schemes and supports sub-object safety by narrowing object bounds. 
}

\ignore{
\subsection{LLVM Aliasing Analysis}
A memory object used in LLVM Aliasing Analysis\cite{llvmaa}  is a region of a memory space reserved by a memory allocation like alloca for stack, allocation calls for heap, and global variable definition. It is expressed as a pair of its starting address and a static size. Based on this information, LLVM Aliasing Analysis provides the method to determine whether or not pointers in the code reference or modify the same memory object. LLVM Alaising Analysis is working well with accessing certain address with the same pointer, However, when the pointer name is changed from assigning the pointer to another pointer, it usually fails to recognize that those pointers are pointing the same memory object. Our taint analysis can track the pointer after assigning the pointer to another pointer.
}

%% file: Tex/Z.Conclusion.tex
\section{Conclusions}

We proposed \mname{}, a RISC-V capability architecture for full memory safety. For capability enforcement, we proposed DPT, a generalized data-pointer tagging method. To achieve optimal performance, we investigated lightweight hardware extensions for DPT based on the RISC-V BOOM core. \mname{} achieved full memory safety at a 10.8\% average slowdown across the SPEC 2017 C/C++ workloads. 
